\documentclass[
reprint,
superscriptaddress,
amsmath,amssymb,
aps,
pra,
floatfix,
]{revtex4-2}

\usepackage{graphicx}
\usepackage{bm}
\usepackage{float} 
\usepackage[colorlinks=true,allcolors=blue]{hyperref}
\usepackage{xcolor}
\usepackage{multirow}
\usepackage[T1]{fontenc}
\usepackage{newtxtext,newtxmath}
\usepackage{booktabs}
\usepackage{soul}

\begin{document}

\title{Improved systematic evaluation of a strontium optical clock with uncertainty below $1\times 10^{-18}$}

\author{Zhi-Peng~{Jia}$^{\#}$}
\affiliation{Hefei National Research Center for Physical Sciences at the Microscale and School of Physical Sciences, University of Science and Technology of China, Hefei 230026, China}
\affiliation{Shanghai Research Center for Quantum Sciences and CAS Center for Excellence in Quantum Information and Quantum Physics, University of Science and Technology of China, Shanghai 201315, China}

\author{Jie~{Li}$^{\#}$}
\affiliation{Hefei National Research Center for Physical Sciences at the Microscale and School of Physical Sciences, University of Science and Technology of China, Hefei 230026, China}
\affiliation{Shanghai Research Center for Quantum Sciences and CAS Center for Excellence in Quantum Information and Quantum Physics, University of Science and Technology of China, Shanghai 201315, China}

\author{De-Quan~{Kong}$^{\#}$}
\affiliation{Hefei National Research Center for Physical Sciences at the Microscale and School of Physical Sciences, University of Science and Technology of China, Hefei 230026, China}
\affiliation{Shanghai Research Center for Quantum Sciences and CAS Center for Excellence in Quantum Information and Quantum Physics, University of Science and Technology of China, Shanghai 201315, China}

\author{Xiang~{Zhang}$^{\#}$}
\affiliation{Hefei National Research Center for Physical Sciences at the Microscale and School of Physical Sciences, University of Science and Technology of China, Hefei 230026, China}
\affiliation{Shanghai Research Center for Quantum Sciences and CAS Center for Excellence in Quantum Information and Quantum Physics, University of Science and Technology of China, Shanghai 201315, China}

\author{Hai-Wei~{Yu}}
\affiliation{Hefei National Research Center for Physical Sciences at the Microscale and School of Physical Sciences, University of Science and Technology of China, Hefei 230026, China}
\affiliation{Shanghai Research Center for Quantum Sciences and CAS Center for Excellence in Quantum Information and Quantum Physics, University of Science and Technology of China, Shanghai 201315, China}

\author{Xiao-Yong~{Liu}}
\affiliation{Hefei National Research Center for Physical Sciences at the Microscale and School of Physical Sciences, University of Science and Technology of China, Hefei 230026, China}
\affiliation{Shanghai Research Center for Quantum Sciences and CAS Center for Excellence in Quantum Information and Quantum Physics, University of Science and Technology of China, Shanghai 201315, China}

\author{Yu-Chen~{Zhang}}
\affiliation{Hefei National Research Center for Physical Sciences at the Microscale and School of Physical Sciences, University of Science and Technology of China, Hefei 230026, China}
\affiliation{Shanghai Research Center for Quantum Sciences and CAS Center for Excellence in Quantum Information and Quantum Physics, University of Science and Technology of China, Shanghai 201315, China}

\author{Yuan-Bo~{Wang}}
\affiliation{Hefei National Research Center for Physical Sciences at the Microscale and School of Physical Sciences, University of Science and Technology of China, Hefei 230026, China}
\affiliation{Shanghai Research Center for Quantum Sciences and CAS Center for Excellence in Quantum Information and Quantum Physics, University of Science and Technology of China, Shanghai 201315, China}

\author{Xian-Qing~{Zhu}}
\affiliation{Hefei National Research Center for Physical Sciences at the Microscale and School of Physical Sciences, University of Science and Technology of China, Hefei 230026, China}
\affiliation{Shanghai Research Center for Quantum Sciences and CAS Center for Excellence in Quantum Information and Quantum Physics, University of Science and Technology of China, Shanghai 201315, China}

\author{Jia-Hao~{Zhang}}
\affiliation{Hefei National Research Center for Physical Sciences at the Microscale and School of Physical Sciences, University of Science and Technology of China, Hefei 230026, China}
\affiliation{Shanghai Research Center for Quantum Sciences and CAS Center for Excellence in Quantum Information and Quantum Physics, University of Science and Technology of China, Shanghai 201315, China}

\author{Ming-Yi~{Zhu}}
\affiliation{Hefei National Research Center for Physical Sciences at the Microscale and School of Physical Sciences, University of Science and Technology of China, Hefei 230026, China}
\affiliation{Shanghai Research Center for Quantum Sciences and CAS Center for Excellence in Quantum Information and Quantum Physics, University of Science and Technology of China, Shanghai 201315, China}

\author{Pei-Jun~{Feng}}
\affiliation{Hefei National Laboratory, University of Science and Technology of China, Hefei 230088, China}

\author{Xing-Yang~{Cui}}
\affiliation{Hefei National Research Center for Physical Sciences at the Microscale and School of Physical Sciences, University of Science and Technology of China, Hefei 230026, China}
\affiliation{Shanghai Research Center for Quantum Sciences and CAS Center for Excellence in Quantum Information and Quantum Physics, University of Science and Technology of China, Shanghai 201315, China}
\affiliation{Hefei National Laboratory, University of Science and Technology of China, Hefei 230088, China}

\author{Ping~{Xu}}
\affiliation{Hefei National Research Center for Physical Sciences at the Microscale and School of Physical Sciences, University of Science and Technology of China, Hefei 230026, China}
\affiliation{Shanghai Research Center for Quantum Sciences and CAS Center for Excellence in Quantum Information and Quantum Physics, University of Science and Technology of China, Shanghai 201315, China}
\affiliation{Hefei National Laboratory, University of Science and Technology of China, Hefei 230088, China}

\author{Xiao~{Jiang}}
\affiliation{Hefei National Research Center for Physical Sciences at the Microscale and School of Physical Sciences, University of Science and Technology of China, Hefei 230026, China}
\affiliation{Shanghai Research Center for Quantum Sciences and CAS Center for Excellence in Quantum Information and Quantum Physics, University of Science and Technology of China, Shanghai 201315, China}
\affiliation{Hefei National Laboratory, University of Science and Technology of China, Hefei 230088, China}

\author{Xiang-Pei~{Liu}}
\affiliation{Hefei National Research Center for Physical Sciences at the Microscale and School of Physical Sciences, University of Science and Technology of China, Hefei 230026, China}
\affiliation{Shanghai Research Center for Quantum Sciences and CAS Center for Excellence in Quantum Information and Quantum Physics, University of Science and Technology of China, Shanghai 201315, China}
\affiliation{Hefei National Laboratory, University of Science and Technology of China, Hefei 230088, China}

\author{Peng~{Liu}}
\affiliation{Shanghai Research Center for Quantum Sciences and CAS Center for Excellence in Quantum Information and Quantum Physics, University of Science and Technology of China, Shanghai 201315, China}
\affiliation{Hefei National Laboratory, University of Science and Technology of China, Hefei 230088, China}

\author{Han-Ning~{Dai}}
\email{daihan@ustc.edu.cn}
\affiliation{Hefei National Research Center for Physical Sciences at the Microscale and School of Physical Sciences, University of Science and Technology of China, Hefei 230026, China}
\affiliation{Shanghai Research Center for Quantum Sciences and CAS Center for Excellence in Quantum Information and Quantum Physics, University of Science and Technology of China, Shanghai 201315, China}
\affiliation{Hefei National Laboratory, University of Science and Technology of China, Hefei 230088, China}

\author{Yu-Ao~{Chen}}
\email{yuaochen@ustc.edu.cn}
\affiliation{Hefei National Research Center for Physical Sciences at the Microscale and School of Physical Sciences, University of Science and Technology of China, Hefei 230026, China}
\affiliation{Shanghai Research Center for Quantum Sciences and CAS Center for Excellence in Quantum Information and Quantum Physics, University of Science and Technology of China, Shanghai 201315, China}
\affiliation{Hefei National Laboratory, University of Science and Technology of China, Hefei 230088, China}
\affiliation{New Cornerstone Science Laboratory, School of Emergent Technology, University of Science and Technology of China, Hefei 230026, China}

\author{Jian-Wei~{Pan}}
\email{pan@ustc.edu.cn}
\affiliation{Hefei National Research Center for Physical Sciences at the Microscale and School of Physical Sciences, University of Science and Technology of China, Hefei 230026, China}
\affiliation{Shanghai Research Center for Quantum Sciences and CAS Center for Excellence in Quantum Information and Quantum Physics, University of Science and Technology of China, Shanghai 201315, China}
\affiliation{Hefei National Laboratory, University of Science and Technology of China, Hefei 230088, China}

\date{\today}
\makeatletter\let\thetitle\@title\makeatother

\begin{abstract}
\noindent We report a systematic uncertainty of $9.2\times 10^{-19}$ for the USTC Sr1 optical lattice clock, achieving accuracy at the level required for the roadmap of the redefinition of the SI second. A finite-element model with {\it in situ}-validated, spatially-resolved chamber emissivity reduced blackbody radiation shift uncertainty to $6.3\times 10^{-19}$. 
Concurrently, an externally mounted lattice cavity combined with a larger beam waist suppressed density shifts. 
Enhanced lattice depth modulation consolidated lattice light shift uncertainty to $6.3\times 10^{-19}$ by enabling simultaneous determination of key polarizabilities and magic wavelength. 
Magnetic shifts were resolved below $10^{-18}$ via precision characterization of the second-order Zeeman coefficient. 
Supported by a crystalline-coated ultra-low-expansion cavity-stabilized laser and refined temperature control suppressing BBR fluctuations, the clock also achieves a frequency stability better than $1\times10^{-18}$ at 30,000-s averaging time. These developments collectively establish a new benchmark in USTC Sr1 clock performance and pave the way for high-accuracy applications in metrology and fundamental physics.
\end{abstract} 

\maketitle

\section{Introduction}~\label{sec.1}

Optical lattice clocks \cite{Ludlow2015,McGrew2018,Takamoto2020,Aeppli2024} and trapped-ion clocks \cite{Marshall2025,Zhang2023,Tofful2024,Hausser2025,Zhang2025} have revolutionized precision metrology, emerging as prime candidates for redefining the SI second~\cite{Dimarcq2024}. These systems employ ultra-stable lasers interrogating ultranarrow atomic transitions, with accuracy fundamentally limited by systematic uncertainties from environmental perturbations. Among multiple prerequisites for SI second redefinition, a key requirement is to deploy at least three optical clocks based on the same reference transition at different institutions, each with a comprehensive uncertainty budget $\leq2\times10^{-18}$.
Although this requirement has not yet been fully met, several systems have recently achieved the uncertainty threshold. For ${}^{87}\text{Sr}$ clocks, two independent teams have demonstrated uncertainties below $2\times10^{-18}$~\cite{Aeppli2024,Lu2025}. Similarly, two ${}^{27}\text{Al}^+$ clock teams have also reached this level~\cite{Ma2024,Marshall2025}. 
Closing this gap necessitates deploying additional clocks with sub-$2\times10^{-18}$ uncertainties.

Based on the 698~nm clock transition, one of the secondary representations of the second (SRS), strontium optical lattice clocks currently facilitate the most precise metrology experiments. Their accuracy is constrained by systematic perturbations including external field interactions and interatomic collisions~\cite{Ludlow2006,Bothwell2019}, with blackbody radiation (BBR) and lattice light shifts dominating the uncertainty budget~\cite{Westergaard2011,Safronova2013,Brown2017,Ushijima2018,Nemitz2019,Lisdat2021,Kim2023}.
For BBR shifts, while the dynamic term is well-characterized through empirically constrained atomic models~\cite{Lisdat2021}, uncertainty reduction demands precise knowledge of the radiation field temperature. 
Although BBR shields improve thermal uniformity~\cite{Zhang2021,Zhang2022} and cryogenic environments suppress the radiation field~\cite{Hassan2025,Zhang2025}, accurate in-vacuum temperature determination remains crucial~\cite{Jin2023}. This is typically achieved through direct probe measurements~\cite{Nicholson2015,Bothwell2019} or validated thermal models~\cite{McGrew2018,Lu2022}.
The lattice light shift presents additional complexity, requiring characterization of multiple parameters ranging from empirical correlations~\cite{Brown2017,Bothwell2019} to \textit{ab initio} approaches based on fundamental atomic properties~\cite{Ushijima2018,Nemitz2019}. To suppress lattice light shift uncertainty to $1.0\times10^{-18}$ and below, evaluations based on microscopic perspectives---providing interpretations independent of experimental configurations---are essential.

\begin{figure*}[t]
\centering
\includegraphics[width=0.85\linewidth]{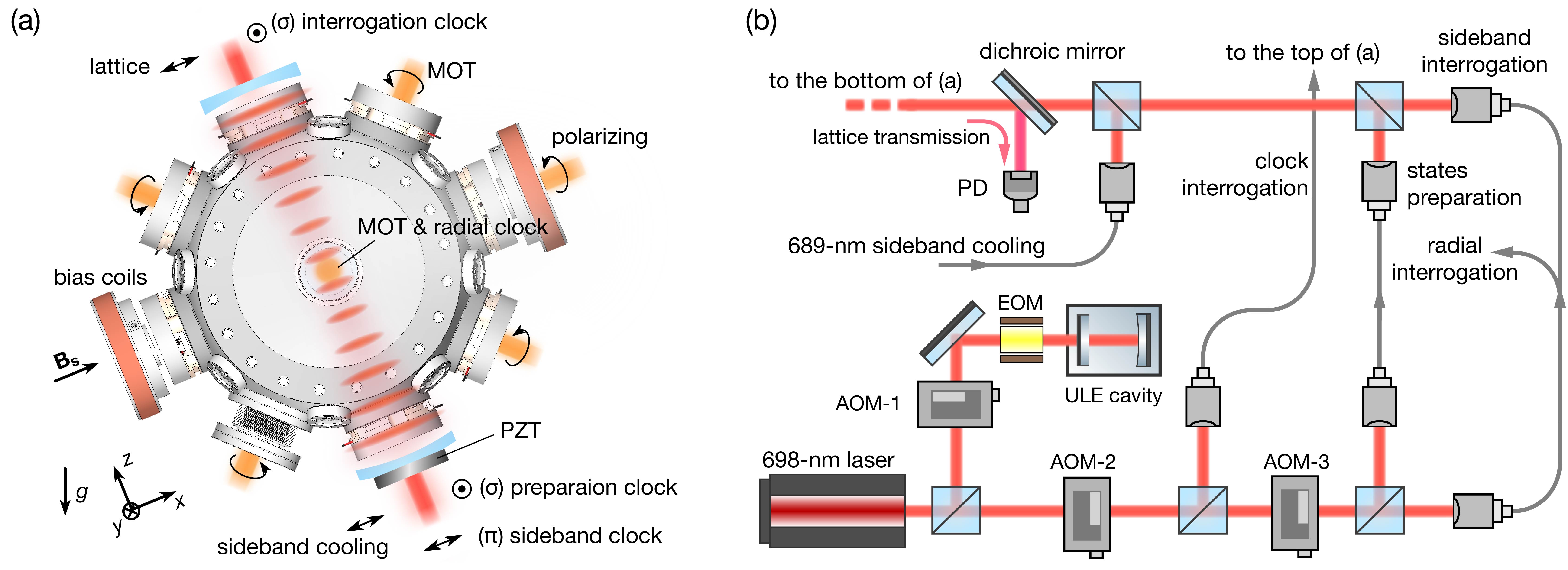}
\caption{(a) Science chamber and lattice cavity configuration. Three orthogonal laser pairs form a MOT: two pairs propagate in the $xOz$ plane, one along $y$. The optical lattice cavity (813 nm HR / 698 nm HT) comprises a biconcave lens and a meniscus lens. Four clock lasers shown in (a): (i) $\sigma$-polarized primary interrogation beam delivered via fiber to the chamber top; (ii) $\sigma$-polarized spin state preparation and (iii) $\uppi$-polarized sideband interrogation beams combined with a 689-nm sideband cooling beam, entering upwards through the bottom viewport; (iv) radial interrogation beam for $T_r$ measurement was delivered through the center viewport. (b) Clock laser system configuration. A 698-nm diode laser PDH-locked to the ULE cavity splits into three beams providing the beams listed in (a). AOM-1 is used to compensate ultrastable laser frequency drifting with AOM-2 for clock feedback. AOM-3 provides frequency hoppings for spin state preparation and sideband interrogation.
\label{fig_setup}}
\end{figure*}

Demonstrating substantial advances beyond our prior evaluation~\cite{Li2024}, this work achieves a systematic uncertainty of $9.2\times10^{-19}$ for the USTC Sr1 strontium optical clock---meeting one key threshold ($2\times10^{-18}$) required as part of the multi-prerequisite roadmap for SI second redefinition. We address the BBR shift through a comprehensive finite-element model incorporating experimental parameters for all vacuum chamber surfaces. 
This framework predicts atomic-position radiation temperatures using surrounding material temperatures and exchange factors, validated against in-vacuum probe data from an isomorphic test chamber and refined by quantifying spatial temperature non-uniformity.
Combined with precision temperature control, this approach achieves a BBR shift uncertainty of $6.3\times10^{-19}$.
For lattice light shift evaluation, we implemented an externally mounted cavity lattice providing 60$\times$ power enhancement and depths exceeding $800E_{\mathrm{r}}$. The high degree of polarization purity in the linearly polarized lattice minimizes vector light shifts, thereby facilitating precise calibration of the tensor shift component. Through strategic modulation protocols in self-comparison measurements, we extracted four critical parameters: electric dipole ($E1$) polarizability derivative, magic frequency, electric quadrupole-magnetic dipole ($E2\text{-}M1$) polarizability, and hyperpolarizability, resulting in a consolidated lattice shift uncertainty of $6.3\times10^{-19}$.
Concurrently, an expanded beam waist suppresses atomic collisions, establishing density shift uncertainty at the $10^{-20}$ level while significantly reducing coupling between density effects and lattice shift evaluations. The second-order Zeeman shift for the magnetically insensitive clock transition is also characterized with $10^{-19}$-level uncertainty. Integration of these precisely evaluated shifts establishes Sr1's total systematic uncertainty below $1\times10^{-18}$, satisfying the accuracy requirement for imminent SI-second redefinition.


\section{Experimental Setup}~\label{sec.2}

The USTC Sr1 optical lattice clock employs a strontium atomic beam generated at $350^\circ\mathrm{C}$, extensively studied in prior work~\cite{Li2023}. Doppler cooling is applied to atoms via the $(5s^2){}^1S_0\!\leftrightarrow\!(5s5p){}^1P_1$ transition at 461 nm. Sequential deceleration through a Zeeman slower precedes deflection and collimation in a 2D-MOT before entry into the science chamber [Fig.~\ref{fig_setup}(a)]. Within this chamber, three orthogonal 461-nm laser pairs confine atoms in a blue magneto-optical trap (MOT), achieving mK-level temperatures. Frequency-swept 689-nm beams then drive $(5s^2){}^1S_0\!\leftrightarrow\!(5s5p){}^3P_1$ transitions: $F'=\frac{11}{2}$ for trapping and $F'=\frac{9}{2}$ for stirring, transferring atoms to a red MOT. This yields $\mu$K-level temperatures and enables efficient optical lattice loading at $U_0=200E_\mathrm{r}$. Key upgrades implemented in this work include: (1) a vacuum-externally mounted optical lattice cavity, (2) enhanced in-lattice cooling protocols, (3) utilization of the magnetically insensitive transition for clock interrogation.

\subsection{Externally Mounted Cavity Lattice}

Precision interrogation of the $^{87}\mathrm{Sr}$ clock transition requires confining atoms in a one-dimensional optical lattice within the Lamb-Dicke regime, suppressing motion-induced Doppler and photon recoil shifts~\cite{Katori2011}. As shown in Fig.~\ref{fig_setup}(a), a vacuum-external optical lattice cavity~\cite{Brown2017,Wang2024} operating at 813 nm provides $60\times$ intensity enhancement. This configuration expands the lattice beam waist from $37~\upmu\mathrm{m}$ to $155~\upmu\mathrm{m}$ compared to the previous reflective setup, requiring four times less input power. The output mirror (bottom) incorporates a piezo ring chip for active cavity length stabilization.

A Pound-Drever-Hall (PDH) lock maintains lattice cavity resonance by stabilizing the fundamental $\mathrm{TEM}_{00}$ mode, where high-frequency feedback modulates laser current and low-frequency components drive the piezo. Simultaneously, a separate PDH loop locks the laser to a 10-cm ULE cavity with finesse $1\times10^{4}$ and long-term drift below $1~\mathrm{kHz/day}$. This dual-loop architecture ensures both relative intensity noise (RIN) suppression and frequency stability.

Background lattice light shifts~\cite{Fasano2021} are suppressed below $1\times10^{-20}$ using a spectrally purified sum-frequency-generation lattice laser combined with volume Bragg grating spectral filtering~\cite{Jia2025}, leveraging the cavity's narrow transmission bandwidth. Polarization characterization for transmitted light yields a linear polarization extinction ratio exceeding $3.7\times10^{4}$.

\subsection{In-lattice Cooling}

Three collinear clock laser beams (two $+z$, one $-z$) overlap with the lattice for atomic interrogation [Fig.~\ref{fig_setup}(a)], enabling spin state preparation, clock operation and sideband spectroscopy. A radially aligned beam ($y$-direction) scans transverse Doppler profiles. Under a 60 mG bias magnetic field (gradient disabled), lattice-trapped atoms undergo simultaneous axial and radial cooling. Axial sideband cooling employs a 689-nm laser spatially overlapped with the lattice to drive the trapping transition~\cite{Nemitz2016,phdthesis2014}, achieving $n_z=0$ population of 0.99 after 50 ms. Concurrent radial cooling via three-axis 689-nm trapping beams~\cite{Ushijima2018} operates for 40 ms with optimized detuning and power, minimizing $T_r$ while preserving axial ground-state purity.

Further enhanced cooling is performed by the energy filtering process~\cite{Falke2014}, where the lattice depth undergoes sigmoidal ramping from $200E_\mathrm{r}$ to $32E_\mathrm{r}$ over 35 ms, holds for 35 ms, then ramps to operational depth. This protocol retains $\sim$10\% atomic population after prior cooling stages. The full preparation sequence delivers 1000 atoms within 700 ms for sustained clock operation.

\begin{figure}[t]
\centering
\includegraphics[width=1.0\linewidth]{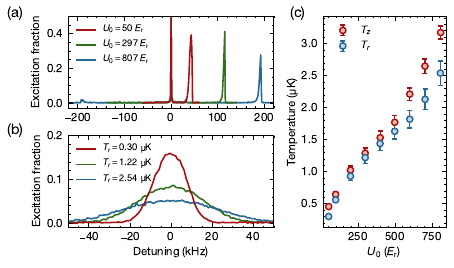}
\caption{
(a) Axial sideband spectra and (b) corresponding radial Doppler profiles at varying lattice depths $U_0$. All spectra recorded after successive axial sideband cooling, radial Doppler cooling, and $32~E_\mathrm{r}$ energy filtering. Increased red sideband amplitudes (indicating axial heating) and broadened radial profiles (reflecting radial heating) show with deeper lattices. (c) Spectral fits yield $T_z$ and $T_r$ values with quasi-linear scaling for $U_0\leqslant 807~E_\mathrm{r}$.
}~\label{fig_sideband_and_radial}
\end{figure}

At the lattice depths below $50E_\mathrm{r}$ [see Fig.~\ref{fig_sideband_and_radial}(c)], the measured axial and radial temperature, $T_z$ and $T_r$, are lower than $0.46~\upmu\mathrm{K}$ and $0.30~\upmu\mathrm{K}$, respectively. When ramped to depths up to $807E_\mathrm{r}$, $T_z$ and $T_r$ scale quasi-linearly to $3.2(1)~\upmu\mathrm{K}$ and $2.5(2)~\upmu\mathrm{K}$, with axial excitation attributed to parametric heating from lattice light RIN.

\begin{figure}[t]
\centering
\includegraphics[width=\linewidth]{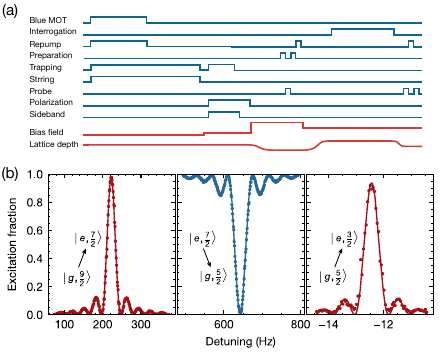}
\caption{
(a) Operational timing sequence for Sr1. The bias magnetic field starts at zero with lattice depth initialized at $200~E_\mathrm{r}$. (b) Left/Middle panels: Rabi lineshapes for clock transitions $|g,\frac{9}{2}\rangle\!\to\!|e,\frac{7}{2}\rangle$ (red) and $|e,\frac{7}{2}\rangle\!\to\!|g,\frac{5}{2}\rangle$ (blue) measured at $1.12~\mathrm{G}$ bias field. The fitted linewidths are $24.5~\mathrm{Hz}$ ($222~\mathrm{Hz}$ detuning) and $33.1~\mathrm{Hz}$ ($643~\mathrm{Hz}$ detuning), respectively. These transitions serve to prepare $|g,\frac{5}{2}\rangle$ states. Right panel: Magnetically insensitive clock transition with $0.6~\mathrm{Hz}$ linewidth and $T_\pi=1390$ ms ($-12.4~\mathrm{Hz}$ detuning at $0.57~\mathrm{G}$ bias field) used for interrogation.
}~\label{fig_timing_and_Rabi_lineshapes}
\end{figure}

\subsection{Magnetically Insensitive Clock States Preparation}

To suppress first-order Zeeman shifts, the Sr1 clock employs magnetically insensitive $\sigma^\mp$-transitions $|g,\pm \frac{5}{2}\rangle\!\leftrightarrow\!|e,\pm\frac{3}{2}\rangle$~\cite{Oelker2019}, where $e$ and $g$ represent the excited and ground clock states, respectively. These transitions exhibit a factor of 22.4 reduction in first-order Zeeman coefficient compared to conventional $\pi$-transitions between stretched states, significantly suppressing magnetic noise coupling to clock frequency fluctuations.

Fig.~\ref{fig_setup}(a) shows the clock laser beams integrated with lattice laser optics at the science chamber, while beam preparation stages including splitting and modulation appear in Fig.~\ref{fig_setup}(b). State preparation starts simultaneously with sideband and radial cooling [Fig.~\ref{fig_timing_and_Rabi_lineshapes}(a)], where circularly polarized 689-nm light drives stirring transitions to optically pump atoms into $|g,\pm\frac{9}{2}\rangle$ stretched states. During lattice ramp-down to $32E_\mathrm{r}$ for energy filtering, two successive circularly polarized clock pulses transfer atoms from $|g,\pm\frac{9}{2}\rangle$ to $|e,\pm\frac{7}{2}\rangle$ and finally to $|g,\pm\frac{5}{2}\rangle$ states. Magic frequencies for both transitions are independently calibrated, enabling active light shift compensation below $0.1~\mathrm{Hz}$, negligible compared to spectral widths in Fig.~\ref{fig_timing_and_Rabi_lineshapes}(b). This protocol achieves state purity exceeding $98.5\%$.

\subsection{Clock Stability}

The Sr1 clock employs Rabi interrogation using a crystalline-coated room-temperature ultra-low-expansion (ULE) cavity-stabilized laser \cite{Zhu2024} with $\uppi$-pulse duration of 1390 ms. During operation, an acousto-optic modulator (AOM) alternately shifts the clock laser frequency to interrogate both $\sigma^\pm$ magnetically insensitive transitions at opposite Rabi lineshape slopes. Excitation fractions are measured via 461-nm fluorescence detection, with a digital  servo locking the clock laser frequency to the atomic transition center.

\begin{figure}[t]
\centering
\includegraphics[width=\linewidth]{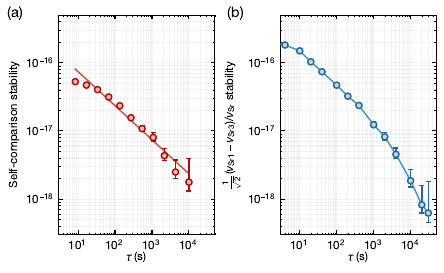}
\caption{
(a) Self-comparison stability of the Sr1 clock with $T_\pi=1390~\text{ms}$. The fitted stability $2.3\times10^{-16}/\sqrt{\tau}$ follows Dick-effect-limited scaling. (b) The frequency stability of the Sr1 clock, derived from the Allan deviation of its comparison with an independent clock Sr3 divided by $\sqrt{2}$, reaches $6.2\times10^{-19}$ at an averaging time of $\tau = 30,000$ s.
}~\label{fig_stability}
\end{figure}

The system achieves fractional frequency stability of $2.3\times10^{-16}/\sqrt{\tau}$ in self-comparisons [Fig.~\ref{fig_stability}(a)], enabling systematic shift evaluation at the low-$10^{-18}$ level. Measured by comparing against an independent Sr3 clock, Sr1's frequency stability reaches $6.2\times10^{-19}$ at 30,000-s averaging time [Fig.~\ref{fig_stability}(b)], indicating systematic drift below the total uncertainty threshold.

\section{Systematic Evaluations}~\label{sec.3}

Systematic effects perturbing the clock transition must be characterized to determine the unperturbed frequency critical for optical clock accuracy. When couplings between systematic effects and the clock transition remain stable, active modulation of perturbation sources enables direct measurement of induced frequency shifts. Combining these measurements with precise theoretical models allows determination of both systematic frequency shifts and their associated uncertainties.

This section details our evaluation approach. First, we establish a blackbody radiation (BBR) shift model based on atomic-position equivalent radiation temperature, using dynamic and static polarizabilities to quantify shifts and uncertainties. To extract model parameters to derive operational shifts and uncertainties, we employ active modulation techniques for three dominant perturbations, atomic collisions, lattice light shifts, and Zeeman effects. Finally, we incorporate analyses of minor systematic effects to present the complete uncertainty budget for USTC Sr1 in Table~\ref{tab_total_uncertainty}.

\begin{figure}[t]
\centering
\includegraphics[width=\linewidth]{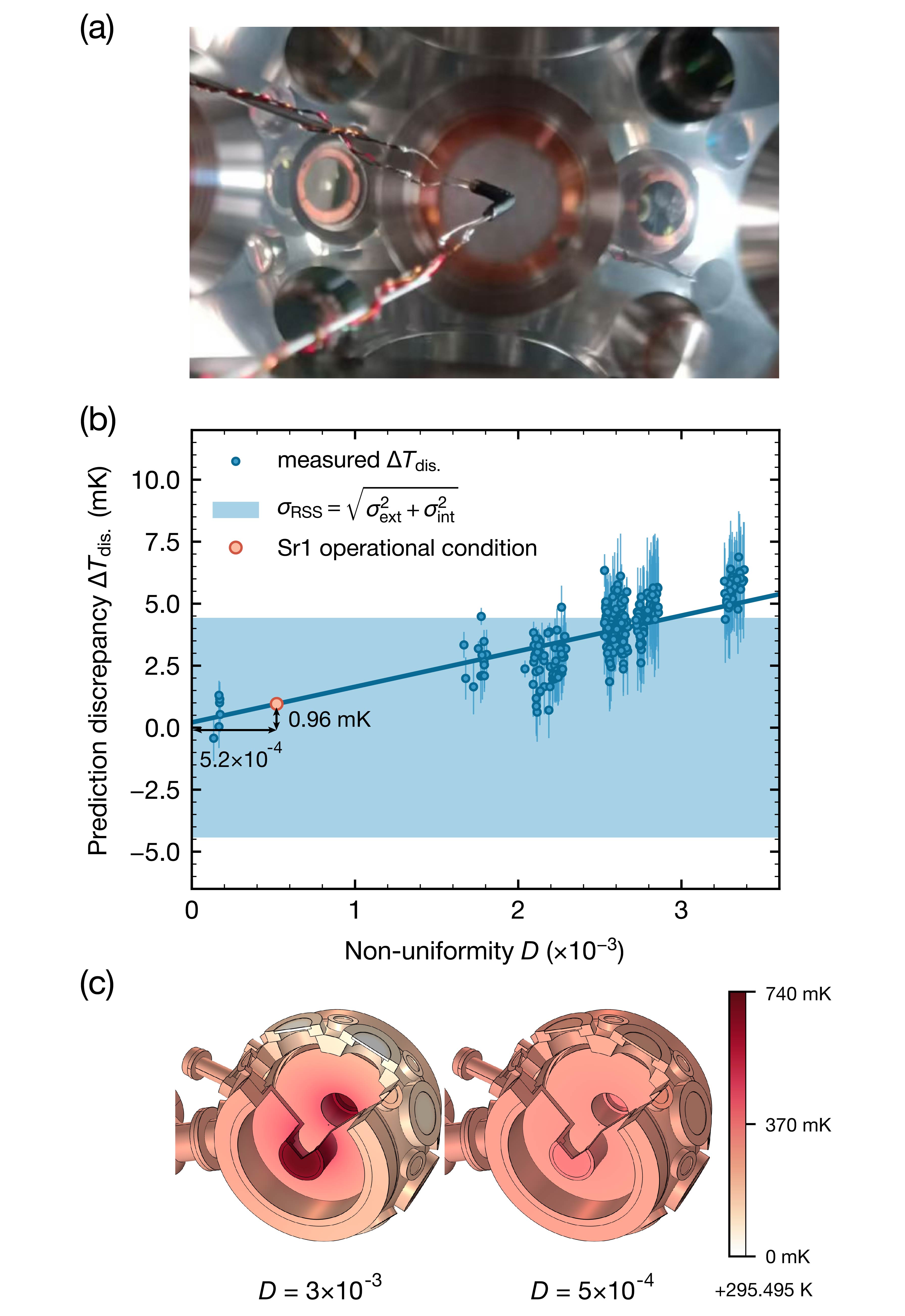}
\caption{(a) Configuration of in-vacuum temperature probes within the isomorphic test chamber. (b) FE-prediction discrepancy $\Delta T_\mathrm{dis.} = T_\mathrm{a,pred.} - T_\mathrm{in\text{-}vac.}$ versus non-uniformity $D$ for the test chamber. Each data point represents a 30-minute comparison. The linear scaling of the discrepancy with $D$ is indicated by a fit ($\chi_\mathrm{red}^2 = 0.9$). The light blue band represents the $\pm1\sigma$ uncertainty ($\sigma_\mathrm{RSS}$), and the red circle ($D = 5.2\times10^{-4}$) marks the typical operational condition of the Sr1 science chamber. (c) Steady-state temperature distributions of the science chamber from FE simulations. Left (right) panel exhibits a non-uniformity of $D = 3\times10^{-3}$ ($D = 5\times10^{-4}$) with a peak-to-peak temperature range of 740 mK (123 mK).
}
\label{fig_BBR-1}
\end{figure}

\subsection{Blackbody Radiation Shift}

The blackbody radiation (BBR) shift represents the most significant systematic effect in strontium optical lattice clocks. Its frequency shift is expressed as~\cite{Lisdat2021}:
\begin{equation}
	\Delta \nu_{\mathrm{BBR}} = -\frac{\Delta \alpha_{\mathrm{stat}}}{2 h}\left\langle E^2\right\rangle_{T_{\mathrm{a}}}\left[1+\eta\left(T_{\mathrm{a}}\right)\right] = \Delta\nu_{\mathrm{stat}}\left(T_{\mathrm{a}}\right) + \Delta\nu_{\mathrm{dyn}}\left(T_{\mathrm{a}}\right),
	\label{eq_BBR_shift}
\end{equation}
where $\Delta \alpha_{\mathrm{stat}}$ is the differential static polarizability and $\langle E^2\rangle_{T_{\mathrm{a}}} = (8.319430\pm15)~\mathrm{(V/cm)^2}(T_\mathrm{a}/300\mathrm{K})^4$ is the mean-square electric field at atomic position temperature $T_\mathrm{a}$, with $\eta$ denoting the dynamic correction. The shift includes a static component $\Delta\nu_{\mathrm{stat}}$ scaling with $T_\mathrm{a}^4$ and a dynamic component $\Delta\nu_{\mathrm{dyn}}$ with higher-order temperature dependencies. 

\textit{Finite-Element Model} ---
We determine $T_\mathrm{a}$ using a finite-element (FE) model of the science chamber. Chamber surfaces are partitioned into segments with assigned parameters: emissivity ($\epsilon$), specular reflectivity ($R_\mathrm{s}$), diffuse reflectivity ($R_\mathrm{d}$), and transmissivity ($T$), satisfying $\epsilon + R_\mathrm{s} + R_\mathrm{d} + T = 1$. Due to high FE model sensitivity to these parameters, all vacuum components were calibrated at the National Synchrotron Radiation Laboratory (NSRL) to ensure reliability, with measured values incorporated into the model~\cite{Yu2025}.

The steady-state FE solution~\cite{Dolezal2015,Xiong2021} relates $T_\mathrm{a}$ to surrounding surface temperatures $T_i$:
\begin{equation}
	T_{\mathrm{a}}^4 = \sum_{i=1} F_i T_i^4,
	\label{eq_BBR_FE}
\end{equation}
where exchange factors $F_i = \partial T_\mathrm{a}/\partial T_i$ are obtained by perturbing each surface temperature in the FE model. This approach enables real-time $T_\mathrm{a}$ determination from measured $T_i$ without prohibitive computational costs. 
Seventeen resistance temperature detectors (RTDs), each with a calibration uncertainty of $7.3~\mathrm{mK}$, were installed on the science chamber. This configuration enabled the establishment of a FE model [Eq.~\ref{eq_BBR_FE}] with an uncertainty of $\sigma_\mathrm{ext}=1.9~\mathrm{mK}$ in predicting the atomic temperature $T_\mathrm{a}$, as detailed in our previous work~\cite{Yu2025}.

\begin{figure}[t]
\centering
\includegraphics[width=\linewidth]{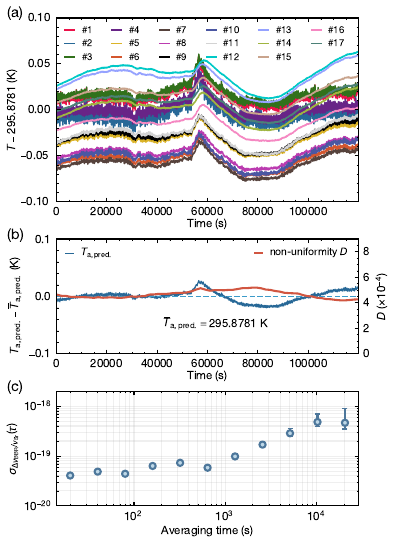}
\caption{
(a) Temporal evolution of all RTD temperatures recorded from the science chamber over 33.3 h. (b) Corresponding FE-predicted temperature $T_\mathrm{a,pred}$ and non-uniformity $D$. 
(c) Allan deviation of the predicted fractional BBR shift remaining $4.8\times10^{-19}$ at an averaging time of 10,000 s and $4.6\times10^{-19}$ at 20,000 s.
}
\label{fig_BBR-2}
\end{figure}

\textit{Model Validation} ---
We validated the FE model using an isomorphic test chamber equipped with two in-vacuum central RTDs [Fig.~\ref{fig_BBR-1}(a)], spaced about 5~mm apart, each calibrated against a standard platinum resistance thermometer (SPRT) with an uncertainty of $5.6~\mathrm{mK}$. The average temperature from these RTDs, $T_\text{in-vac.}=(T_\text{RTD1}+T_\text{RTD2})/2$, was determined with an uncertainty of $\sigma_\mathrm{int}=5.6/\sqrt{2}~\mathrm{mK}=4.0~\mathrm{mK}$. For the science chamber itself, seventeen RTDs with identical geometric configuration and individual calibration uncertainty ($7.3~\mathrm{mK}$, as noted previously) monitored the chamber wall temperatures.

\begin{table}[b]
  \centering
  \caption{
  Uncertainty contributors to the radiation temperature $T_\mathrm{a}$ and uncertainty contributors to the total shift $\Delta\nu_\text{BBR}$.
  }~\label{tab_BBR_shift}
  \begin{tabular}{@{} p{0.4\linewidth} p{0.4\linewidth} @{}} 
  \toprule[1pt]
    Items  &  Value \\
    \midrule[0.5pt]
    $\sigma_\text{int}$  &  4.0 mK \\
    $\sigma_\text{ext}$  &  1.9 mK \\
    ${T_\mathrm{a}}$  &  $295.8781\pm0.0044~\mathrm{K}$ \\
    \midrule[0.5pt]
    $\sigma_{\text{BBR},{T_\mathrm{a}}}/\nu_\text{Sr}$  &  $3.1\times10^{-19}$  \\
    $\sigma_\text{BBR,stat}/\nu_\text{Sr}$  &  $1.3\times10^{-19}$  \\
    $\sigma_\text{BBR,dyn}/\nu_\text{Sr}$  &  $5.4\times10^{-19}$  \\
    \midrule[0.5pt]
    $\Delta\nu_\text{BBR}/\nu_\text{Sr}$  &  $-50222.8\pm6.3\times10^{-19}$  \\
  \bottomrule[1pt]
  \end{tabular}
\end{table}

For an ideal situation where the chamber wall temperature distribution is fully uniform and constant, $T_\mathrm{a}$ equals the chamber temperature. When slight non-uniformity occurs, Eq.~\eqref{eq_BBR_FE} provides a reasonable prediction of $T_\mathrm{a}$. However, prediction deviation increases with significant chamber temperature non-uniformity. By defining the dimensionless non-uniformity $D$ [Fig.~\ref{fig_BBR-1}(c)] as the weighted standard deviation of all $T_i^4$:
\begin{equation}
    D^2=\frac{1}{T_\mathrm{a}^8}\sum_i F_i(T_i^4-T_\mathrm{a}^4)^2,
\end{equation}
we quantitatively explored this effect using the isomorphic test chamber. The dependence of the prediction discrepancy, $\Delta T_\mathrm{dis.}= T_\mathrm{a,pred.}-T_\text{in-vac.}$, on non-uniformity $D$ is shown in Fig.~\ref{fig_BBR-1}(b). Each data point represents a 30-minute comparison and exhibits an average temperature drifting rate lower than $9~\mathrm{mK}/30~\mathrm{min}$ across all 17 RTDs, thus avoiding prediction ambiguity due to unsteady temperature distribution. 
The data confirm that the discrepancy $\Delta T_\mathrm{dis.}$ scales linearly with non-uniformity $D$, a relationship characterized by the fit $\Delta T_\mathrm{dis.}/\mathrm{K}=1.4D+2.1\times10^{-4}$ with a reduced chi-square $\chi_\mathrm{red}^2=0.90$.

For the Sr1 science chamber, owing to the optical table's secondary air conditioning, the average temperature drifting rate of all RTDs remains below $9~\mathrm{mK}/30~\mathrm{min}$. The typical temperature non-uniformity is also below $5.2\times10^{-4}$ [Fig.~\ref{fig_BBR-2}(b)], and thus the prediction discrepancy is expected to be lower than $1.0~\mathrm{mK}$, consistent with the root sum square $\sigma_\mathrm{RSS}=(\sigma_\mathrm{int}^2+\sigma_\mathrm{ext}^2)^{1/2}=4.4~\mathrm{mK}$. Therefore, the FE prediction is validated by the in-vacuum probes, and $\sigma_\mathrm{RSS}$ characterizes the final systematic uncertainty of $T_\mathrm{a}$ based on the above analysis.

\textit{Clock Operation} ---
Infrared thermography of peripheral equipment (e.g., CCDs) identified extraneous heat sources, and these were deactivated during clock operation. Optimized thermal management of the MOT coil water-cooling system limited spatial temperature gradients to below $110~\mathrm{mK}$ over 33.3 hours of continuous operation [Fig.~\ref{fig_BBR-2}(a)]. 
The predicted $T_\mathrm{a}$ exhibited a peak-to-peak variation below $48~\mathrm{mK}$ during this period [Fig.~\ref{fig_BBR-2}(b)], with the Allan deviation of the resulting fractional BBR shift, $\Delta\nu_\mathrm{BBR}/\nu_\mathrm{Sr}$, remaining at $4.6\times10^{-19}$ for a 20,000-s averaging time [Fig.~\ref{fig_BBR-2}(c)]. During clock operation, the BBR shift is actively compensated every 1000 s based on the model prediction, further suppressing the degradation of clock stability from temperature fluctuations.

The radiation temperature was determined to be $T_\mathrm{a} = 295.8781(44)~\mathrm{K}$ (Table~\ref{tab_BBR_shift}). At this temperature, the static shift component was obtained as $\Delta\nu_{\mathrm{stat}} = -2.15571(6)~\mathrm{Hz}$ from the $T_\mathrm{a}^4$ scaling~\cite{Middelmann2012}, and the dynamic shift was calculated as $\Delta\nu_{\mathrm{dyn}} = -140.16(23)~\mathrm{mHz}$ following the method in Ref.~\cite{Lisdat2021,Aeppli2024}. Combining these components yields the total BBR-induced fractional frequency shift of $(-50222.8 \pm 6.3) \times 10^{-19}$.

\subsection{Density Shift}

\begin{figure}[t]
\centering
\includegraphics[width=\linewidth]{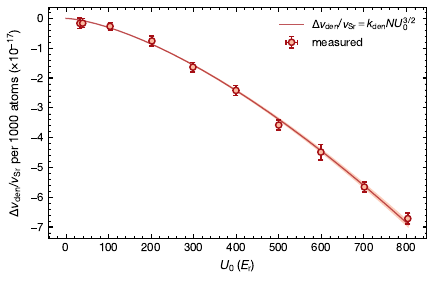}
\caption{
Fractional density shift per 1000 atoms versus lattice depth, verifying Eq.~\eqref{eq_density_shift}. The fit yields $k_\mathrm{den}=-3.03(5)\times10^{-24}E_\mathrm{r}^{-3/2}$ with $\chi_\mathrm{red}^2=0.43$.
}~\label{fig_density_shift}
\end{figure}

Interatomic collisions within lattice sites induce density shifts perturbing the clock transition frequency and distorting Rabi lineshapes~\cite{Bishof2011,Swallows2012}. These shifts originate from residual $s$-wave~\cite{Campbell2009,Swallows2011} and $p$-wave scattering~\cite{Lemke2011,Bishof2011}, modulated by technical factors including clock laser-lattice misalignment~\cite{Zhou2023}, excitation fraction~\cite{Bishof2011,Ludlow2011}, and temperature-depth correlations~\cite{Swallows2012}.
Empirically, the density shift follows~\cite{Nicholson2015,Lu2023,Li2024}:
\begin{equation}
	\Delta\nu_\mathrm{den}/\nu_\mathrm{Sr} = k_\mathrm{den} N_\mathrm{a} U_0^{3/2},
	\label{eq_density_shift}
\end{equation}
where $N_\mathrm{a}$ is the atom number and $k_\mathrm{den}$ the density shift coefficient. We implemented a broad-waist lattice suppressing this effect through reduced atomic density~\cite{Nicholson2015}.

To quantify density shifts, we performed self-comparison measurements applying a \hbox{$\sim\!4\times$} lever-arm atom number modulation across 10 lattice depths ($33$ to $803~E_\mathrm{r}$).
Fig.~\ref{fig_density_shift} shows the density shift per 1000 atoms versus lattice depth. The dependence fitted using Eq.~\eqref{eq_density_shift} yields \hbox{$k_\mathrm{den} = -3.03(5)\times10^{-24}~E_\mathrm{r}^{-3/2}$}. At operational conditions \hbox{$N_\mathrm{a}\!=\!1000(40)$} and \hbox{$U_0^\mathrm{op}\!=\!33.0(1)~E_\mathrm{r}$}, the density shift is \hbox{$\Delta\nu_\mathrm{den}/\nu_\mathrm{Sr}\!=\!-5.7(2)\times10^{-19}$}. This represents a 30-fold reduction compared to previous work, enabling significant decoupling from lattice light shifts during depth-modulated self-comparisons.

\subsection{Lattice Light Shift}

\begin{figure}[t]
\centering
\includegraphics[width=\linewidth]{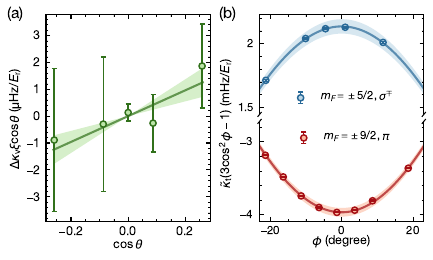}
\caption{
(a) Vector light shift coefficients versus $\cos\theta$ for the $\pi$-transition between stretched states, with linear fit yielding $\Delta\kappa_\mathrm{v}\xi=4.8(2.1)~\upmu\mathrm{Hz}/E_\mathrm{r}$. 
(b) Tensor light shifts versus $\phi$ after common-mode scalar shift subtraction. Data points for stretched-$\pi$ transition (red) and magnetically insensitive transition (blue) are fitted with quadratic functions of $\cos\phi$, yielding ${\kappa}_\mathrm{t}^{e,g}$ values. Shaded regions indicate $\pm1\sigma$ confidence intervals.
}~\label{fig_vector_and_tensor_shift}
\end{figure}

\begin{figure*}[t]
\centering
\includegraphics[width=1\linewidth]{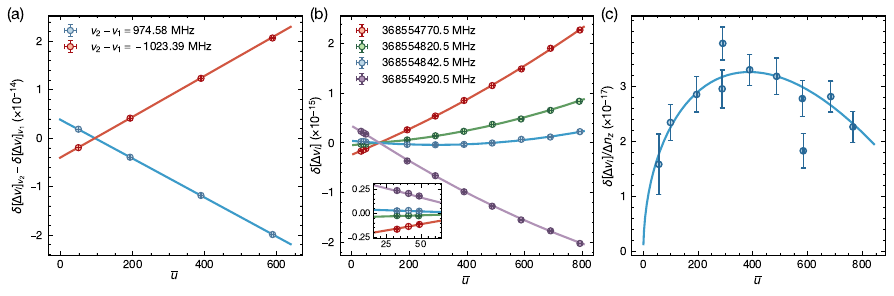}
\caption{(a) Differential self-comparison frequency shifts between lattice frequencies $\nu_2$ and $\nu_1$ within $\pm1~\mathrm{GHz}$ of the magic frequency. A linear fit to $u_z$ (see text) determines $\partial\alpha_{E1}/\partial\nu=1.811(5)\times10^{-11}$ with $\chi_\mathrm{red}^2=1.22$. The abscissa displays $\overline{u}$ for unified visualization.
(b) Lattice light shifts measured at four frequencies near the magic frequency with modulated ($\overline{u}^\mathrm{mod}$) and reference ($\overline{u}^\mathrm{ref}=94.4$) depths. Global iterative fitting resolved correlated parameters, yielding $\beta=-0.466(13)~\upmu\mathrm{Hz}$ and $\nu^\mathrm{m}_{E1}=368554826.6(5)~\mathrm{MHz}$ with $\chi_\mathrm{red}^2=1.39$.
(c) Self-comparison frequency shifts during $n_z$ modulation (points with error bars) and fitted curve (blue), resolving $\alpha_\mathrm{qm}=-1.057(32)~\mathrm{mHz}$ with $\chi_\mathrm{red}^2=0.69$. The ordinate shows shift ratios to $\langle n_z\rangle$ modulation amplitudes for clarity. 
Data at $\overline{u}=290$ and $584$ show slight deviations from the fitted curve, likely arising from unexpected parametric heating, though random fluctuations may also contribute; additional measurements were therefore performed at $\overline{u}=288$, $581$, and $766$ by more carefully controlling the lattice intensity noise to refine the analysis.
}~\label{fig_lattice_light_shift}
\end{figure*}

Optical lattices confine strontium atoms at standing-wave anti-nodes by minimizing AC Stark shifts. Strontium clocks must operate at the magic frequency where $E1$ polarizabilities of clock states balance, yielding near-zero net shift~\cite{Katori2003,Takamoto2005}. 
For experimental implementation, we first align the bias magnetic field parallel to lattice polarization to suppresses vector light shifts and minimizes tensor shift sensitivity to magnetic field angle fluctuations.
Following recent theoretical advances, we apply the comprehensive higher-order model for lattice light shifts~\cite{Ushijima2018,Kim2023,Bothwell2025}:
\begin{equation}
	\begin{aligned}
		\Delta\nu_l = &\left[\frac{\partial\alpha_{E1}}{\partial\nu}(\nu-\nu_{E1}^\mathrm{m})-\alpha_\mathrm{qm}\right]\left(n_z+\frac{1}{2}\right)\overline{u^{1/2}} \\
		&-\left[\frac{\partial\alpha_{E1}}{\partial\nu}(\nu-\nu^\mathrm{m}_{E1})+\frac{3}{2}\beta\left(n_z^2+n_z+\frac{1}{2}\right)\right]\overline{u} \\
		&+2\beta\left(n_z+\frac{1}{2}\right)\overline{u^{3/2}}-\beta \overline{u^2},
	\end{aligned}
	\label{eq_lattice_light_shift}
\end{equation}
where $\overline{u^j}=u^j[1+jk_\mathrm{B}T_r/(uE_\mathrm{r})]^{-1}$ accounts for radial motion averaging at temperature $T_r$. Here $u=U_0/E_\mathrm{r}$ is dimensionless depth, $n_z$ the axial quantum number, $\nu$ the lattice frequency, and $\nu^\mathrm{m}_{E1}$ the $E1$ magic frequency. Coefficients include $\alpha_{E1}$ ($E1$ polarizability), $\alpha_\mathrm{qm}$ ($E2\text{-}M1$ polarizability), and $\beta$ (hyperpolarizability), all in frequency units.

\begin{table}[b]
  \centering
  \caption{Parameters of lattice light shifts: $E1$ polarizability derivative ($\partial_\nu\alpha_{E1}$), $E1$ magic frequency ($\nu^\mathrm{m}_{E1}$), $E2\text{-}M1$ polarizability ($\alpha_\mathrm{qm}$), and hyperpolarizability ($\beta$), compared with literature values. The $\nu^\mathrm{m}_{E1}$ value from Ref.~\cite{Ushijima2018} corresponds to $\uppi$-transition of stretched states.
  }~\label{tab_lattice_shift}
  \begin{tabular}{cccc}
  \toprule[1pt]
    Quantities & This work & Ref.~\cite{Kim2023} & Ref.~\cite{Ushijima2018}\\
    \midrule[0.5pt]
    $\partial_\nu\alpha_{E1}$ ($10^{-11}$)& 1.811(5) & 1.859(5) & 1.735(13) \\
    $\nu^\mathrm{m}_{E1}$ (MHz) & 368554826.6(5) & 368554825.9(4) & 368554465.1(10)\\
    $\alpha_\mathrm{qm}$ (mHz)  & $-1.057(32)$ & $-1.24(5)$ & $-0.962(40)$\\
    $\beta$ ($\upmu$Hz) & $-0.466(13)$ & $-0.51(4)$ & $-0.461(14)$ \\
  \bottomrule[1pt]
  \end{tabular}
\end{table}

{\it Calibration of $E1$ Vector and Tensor Light Shifts} --- Prior to determining the coefficients in Eq.~\eqref{eq_lattice_light_shift}, the $E1$ vector and tensor light shifts must be calibrated. For the $m_F=m_{e(g)}$ state, these shifts are given by~\cite{Shi2015}:
\begin{equation}
\begin{aligned}
    \Delta\nu_\mathrm{v+t}^{e(g)}=&m_{e(g)}\kappa_\mathrm{v}^{e(g)}\xi\cos\theta U_0
    \\
    &+\kappa_\mathrm{t}^{e(g)}\left[3m^2_{e(g)}-F(F+1)\right](3|\vec{\epsilon}_l\cdot\hat{\mathbf{B}}_\mathrm{s}|^2-1)U_0,
\end{aligned}
\label{eq_vec_and_ten}
\end{equation}
where $\kappa_\mathrm{v}$ and $\kappa_\mathrm{t}$ are the vector and tensor polarizabilities, $\theta$ is the angle between the lattice wave vector and the bias magnetic field direction $\hat{\mathbf{B}}_\mathrm{s}$, and $\vec{\epsilon}_l$ is the complex polarization vector with $\xi$ as the polarization parameter. For ideal linear polarization, $\xi=0$ and $(3|\vec{\epsilon}_l\cdot\hat{\mathbf{B}}_\mathrm{s}|^2-1)=(3\cos^2\phi-1)$, with $\phi$ being the angle between $\vec{\epsilon}_l$ and $\hat{\mathbf{B}}_\mathrm{s}$. As detailed in Section~\ref{sec.2}, the Sr1 apparatus adopts a conventional configuration: the lattice wavevector is perpendicular to the bias magnetic field ($\theta=\uppi/2$), and the linearly lattice polarization aligns parallel to the magnetic field ($\phi=0$).

We first calibrated the $E1$ vector light shifts. By varying the bias magnetic field orientation under lattice depth modulation of $\Delta U_0=450~E_\mathrm{r}$, we measured these shifts through symmetric sublevel transition splitting, as shown in Fig.~\ref{fig_vector_and_tensor_shift}(a). This yielded a minimal vector shift coefficient $\Delta\kappa_\mathrm{v}\xi=4.8(2.1)~\upmu\mathrm{Hz}/E_\mathrm{r}$, confirming pure linear lattice polarization with $|\Delta\kappa_\mathrm{v}|=0.22(5) \mathrm{Hz}/E_\mathrm{r}$~\cite{Shi2015}. 

Angular-dependent tensor light shifts were measured through the average frequency of symmetric sublevel transitions. From fits to the tensor shifts of the stretched-$\pi$ and the magnetically insensitive transition [Fig.~\ref{fig_vector_and_tensor_shift}(b)], we extracted $\kappa_\mathrm{t}^{g}=-6.4(2.7)~\upmu\mathrm{Hz}/E_\mathrm{r}$ and $\kappa_\mathrm{t}^{e}=-61.5(2.4)~\upmu\mathrm{Hz}/E_\mathrm{r}$. All these measurements ensured the magnetic field orientation parallel to lattice polarization with $0.1^\circ$ uncertainty.

\textit{Parameter Determination} ---
Characterization of lattice light shift requires measurement of four parameters  in Eq.~\eqref{eq_lattice_light_shift}: $\partial\alpha_{E1}/\partial\nu$, $\nu^\mathrm{m}_{E1}$, $\alpha_\mathrm{qm}$, and $\beta$. During depth modulation, density shifts varying with depth are subtracted based on the calibrated density shift model [Eq.~\eqref{eq_density_shift}] to isolate pure lattice light shifts.

For $\partial\alpha_{E1}/\partial\nu$, we employed differential self-comparison at two lattice frequencies ($\nu_1$, $\nu_2$):
\begin{equation}
	\delta[\Delta\nu_l]_{\nu_2} - \delta[\Delta\nu_l]_{\nu_1} = -\frac{\partial\alpha_{E1}}{\partial\nu}(\nu_2-\nu_1)\delta[u_z],
\end{equation}
where $\delta[\cdot]$ denotes differences between modulated ($u_z^\mathrm{mod}=49$ to $572$) and reference ($u_z^\mathrm{ref}=89$) conditions, with $u_z=\overline{u}-(n_z+1/2)\overline{u^{1/2}}$. At about $\pm1$ GHz detuning around $\nu_{E1}^\mathrm{m}$, we obtained $\partial\alpha_{E1}/\partial\nu=1.811(5)\times10^{-11}$ [Fig.~\ref{fig_lattice_light_shift}(a)] with a reduced chi-square $\chi^2_\mathrm{red}=1.22$ (throughout this article, all uncertainties were scaled by $(\chi^2_\mathrm{red})^{1/2}$ when $\chi_\mathrm{red}^2>1$). We resolved $\beta$ and $\nu^\mathrm{m}_{E1}$ by modulating $\overline{u}$ from 49 to 792 at four frequencies within 150 MHz of $\nu_{E1}^\mathrm{m}$ [Fig.~\ref{fig_lattice_light_shift}(b)].

Determination of $\alpha_\mathrm{qm}$ required $n_z$ state modulation during self-comparisons. The $n_z=1$ state was prepared through: blue sideband excitation $|g,n_z=0\rangle\!\leftrightarrow\!|e,n_z=1\rangle$ at $60E_\mathrm{r}$; removal of residual $|g,n_z=0\rangle$ atoms via 461-nm light; carrier $\uppi$-pulse transfer $|e,n_z=1\rangle\!\to\!|g,n_z=1\rangle$; and finally this achieved $\langle n_z\rangle=1.02(5)$ with maintained spin purity. Self-comparison results between $n_z=0$ (reference) and $n_z=1$ states appear in Fig.~\ref{fig_lattice_light_shift}(c). Then iterative fitting determined $\beta$, $\nu^\mathrm{m}_{E1}$, and $\alpha_\mathrm{qm}$ values as follows.

We performed global iterative fitting~\cite{Ushijima2018} of all modulation data [depth and frequency in Fig.~\ref{fig_lattice_light_shift}(b), $n_z$ state in (c)] using Eq.~\eqref{eq_lattice_light_shift}, propagating uncertainties in $\partial_\nu\alpha_{E1}$, $U_0$, $T_r$, $\nu$, and $\langle n_z\rangle$. Converged results yield:
$\beta = -0.466(13)~\upmu\mathrm{Hz}$, $\nu^\mathrm{m}_{E1} = 368554826.6(5)~\mathrm{MHz}$, $\alpha_\mathrm{qm} = -1.057(32)~\mathrm{mHz}$.
Table~\ref{tab_lattice_shift} compares these values with literature, noting the $\sim$362 MHz discrepancy in $\nu^\mathrm{m}_{E1}$ relative to Ref.~\cite{Ushijima2018} originates from different magnetic sublevel transitions. The determined $\alpha_\mathrm{qm}$ value is further supported by consistency with Ref.~\cite{Drscher2023} and theoretical predictions from Ref.~\cite{Wu2023}.

\textit{Operational Shift Evaluation} ---
At operational conditions $\nu=\nu_{E1}^\mathrm{m}$, $\langle n_z\rangle=0.00(1)$, $U_0^\mathrm{op}=33.0(1)E_\mathrm{r}$ yielding $\overline{u}^\mathrm{op}=32.4(1)$, the lattice light shift is $\Delta\nu_l/\nu_\mathrm{Sr}=80.5(63)\times10^{-19}$. The $6.3\times10^{-19}$ uncertainty is dominated by the $0.47$ MHz uncertainty in $\nu_{E1}^\mathrm{m}$.

\begin{figure}[t]
\centering
\includegraphics[width=\linewidth]{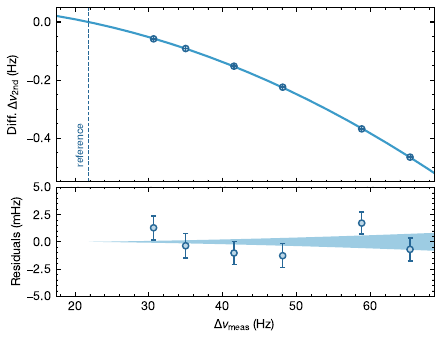}
\caption{
Upper panel: Differential second-order Zeeman shifts between modulated and reference bias magnetic fields. The abscissa $\Delta\nu_\mathrm{meas}$ represents the measured splitting between symmetric magnetically insensitive clock transitions. Blue curve shows fit to Eq.~\eqref{eq_2nd_Zeeman_shift} with vector shifts fixed from independent characterization, yielding $\xi_\mathrm{i}=-0.12277(23)~\mathrm{mHz/Hz^2}$ and $\chi_\mathrm{red}^2=1.42$. Lower panel: Fit residuals (blue error bars) with $\pm1\sigma$ uncertainty (light blue shaded region).
}
\label{fig_2nd_Zeeman_shift}
\end{figure}

\subsection{Zeeman Shift}

Residual first-order effects from magnetic field fluctuations synchronized with clock cycles are bounded by:
\begin{equation}
\Delta\nu_\mathrm{Z1,res} \leqslant \langle S_{2n}-S_{2n-1}\rangle/8,
\end{equation}
where $S_n$ denotes Zeeman splitting between symmetric transitions during the $n$-th cycle. Using symmetric clock transitions between stretched states, this shift for Sr1 is $0\pm8\times10^{-19}$~\cite{Li2024}. Magnetically insensitive clock transitions offer a suppression factor of about $22.4\times$, and thus the uncertainty of $\Delta\nu_\mathrm{Z1,res}$ achieves $0.4\times10^{-19}$.

The second-order Zeeman shift scales quadratically with magnetic field strength and remains unidirectional for all magnetic sublevels~\cite{Shi2015,Bothwell2019}:
\begin{equation}
	\Delta\nu_\mathrm{Z2} = \xi_\mathrm{i} \Delta\nu_\mathrm{Z1}^2,
	\label{eq_2nd_Zeeman_shift}
\end{equation}
where $\xi_\mathrm{i}$ is the second-order coefficient for magnetically insensitive transitions. Here $\Delta\nu_\mathrm{Z1}$ is derived from measured symmetric transition splittings $\Delta\nu_\mathrm{meas}$ as $
\Delta\nu_\mathrm{Z1} = \Delta\nu_\mathrm{meas} - \Delta\nu_\mathrm{v}$, with $\Delta\nu_\mathrm{v}$ representing vector light shifts.

We characterized $\xi_\mathrm{i}$ through self-comparison measurements with a 500-mG reference bias field and 700-1500-mG modulation fields. Vector light shift was independently quantified as $\Delta\nu_\mathrm{v}/U_0 = -7.6(2.8)~\upmu\mathrm{Hz}/E_\mathrm{r}$ for magnetically insensitive transitions. Constraining $\Delta\nu_\mathrm{v}$ in fits to Eq.~\eqref{eq_2nd_Zeeman_shift}, we obtained $\xi_\mathrm{i} = -0.12277(23)~\mathrm{mHz/Hz^2}$ [Fig.~\ref{fig_2nd_Zeeman_shift}].
At operational conditions ($\Delta\nu_\mathrm{Z1} = 20.0~\mathrm{Hz}$, $B^\mathrm{op}_\mathrm{s} \approx 460~\mathrm{mG}$), the second-order Zeeman shift is $\Delta\nu_\mathrm{Z2}/\nu_\mathrm{Sr} = -1144.1(21)\times10^{-19}$.

\subsection{Other Minor Systematic Shifts}

\textit{Residual DC Stark Shift} --- 
This shift originates from static electric fields due to charge accumulation on viewports~\cite{Lodewyck2012}. As reported in our previous work~\cite{Li2024}, the $y$-component of the field [Fig.~\ref{fig_setup}(a)] caused a shift of $1.4(5.2)\times 10^{-21}$. Given $\Delta\nu_\mathrm{DC} \propto E_\mathrm{stat}^2$ scaling and $E_\mathrm{stat} \propto r^{-2}$ dependence, the shorter $y$-axis viewport distance (142 mm vs. 237 mm for $x,z$) would yield an $8\times$ larger shift along $y$. Electrostatic shielding further suppresses $x,z$-axis fields by a factor of 3 according to FE simulations. Combining these, we constrain the total shift to $0.0(0.1)\times 10^{-19}$.

\textit{Background Gas Collisions} --- 
Two-body collisions with residual $H_2$ induce shifts proportional to loss rate. Using the coefficient $-3.0(3)\times10^{-17}$ Hz/(atom/s)~\cite{Alves2019} and measured vacuum lifetime $45(2)$ s, we obtain a shift of $-6.7(0.7)\times 10^{-19}$.

\textit{AOM Phase Chirp} --- 
Thermal transients during AOM activation introduce phase chirp, effectively generating Doppler shifts. In Sr1, the AOM resides within the fiber noise cancellation (FNC) loop, with feedback stabilizing the optical path within $20~\upmu\mathrm{s}$. Using IQ demodulation to extract the instantaneous phase, we calculated the corresponding frequency shift below $1\times 10^{-20}$.

\textit{Line-Pulling Effect} --- 
Residual ground-state atoms in non-interrogated sublevels perturb Rabi lineshapes. For Sr1's $\uppi$-pulse duration ($T_\mathrm{i}^\mathrm{op}=1390$ ms) and $98.5\%$ spin purity, uniform residual population leads to a shift $<1\times10^{-21}$, with the worst case $\leq9\times10^{-21}$. Thus we assign a systematic shift of $0.0(0.1)\times 10^{-19}$.

\textit{Lattice Tunneling} --- 
The lattice geometry, tilted at $22.5^\circ$ relative to gravity [Fig.~\ref{fig_setup}(a)], yields a tunneling-induced frequency shift~\cite{Lemonde2005} below $1\times10^{-20}$ for an interrogation pulse duration $T_\mathrm{i}^\mathrm{op} = 1390~\mathrm{ms}$ at a depth of $U_0^\mathrm{op} = 33.0(1)~E_\mathrm{r}$.

\textit{Probe AC Stark Shift} ---
The clock laser induces a weak ac Stark shift. Using the coefficient from~\cite{Xu2021} and operational intensity ($T_\mathrm{i}^\mathrm{op}=1390$ ms), we calculate a shift magnitude of $-7\times 10^{-21}$. This value is assigned as the shift uncertainty.

\textit{Servo Error} --- 
This shift originates from loop delay $T_s=-0.29(2)$ s and clock laser frequency drift~\cite{Li2024}. Automatic recalibration and compensation every 1000 s suppress the drift rate to $0.0(0.1)$ mHz/s, bounding the servo error to $0.0(0.7)\times 10^{-19}$.

\begin{table}[tb]
  \centering
  \caption{\centering USTC Sr1 uncertainty budget.}
  \begin{tabular}{lcc}
  \toprule[1pt]
    Systematic & Shift ($10^{-19}$) & Uncertainty ($10^{-19}$) \\
    \midrule[0.5pt]
    BBR & $-50222.8$ & $6.3$ \\
    Density & {$-5.7$} & {$0.2$} \\
    Lattice & $80.5$ & $6.3$ \\
    1 st Zeeman & $0$ & $0.4$ \\
    2nd Zeeman & $-1144.1$ & $2.1$\\
    DC Stark & $0$ & {$<0.1$} \\
    Background gas & $-6.7$ & $0.7$\\
    AOM phase chirp & $0$ & $<0.1$\\
    Line pulling & $0$ & {$<0.1$}\\
    Lattice Tunneling & $0$ & {$<0.1$}\\
    Probe AC Stark & $0$ & {$<0.1$}\\
    Servo error & $0$ & $0.7$\\
    \midrule[0.5pt]
    \textbf{Total} & {$\mathbf{-51298.8}$} & $\mathbf{9.2}$ \\
  \bottomrule[1pt]
  \end{tabular}~\label{tab_total_uncertainty}
\end{table}


\section{Summary}~\label{sec.4}

This work reports systematic upgrades to the USTC Sr1 optical lattice clock, achieving a total systematic uncertainty of $9.2\times10^{-19}$. A crystalline-coated ULE cavity provides clock stability of $6.2\times10^{-19}$ at 30,000-s averaging time. Refined blackbody radiation shift modeling, validated by in-vacuum probes and non-uniformity analysis, constrains its uncertainty contribution to $6.3\times10^{-19}$. Atomic density shifts are suppressed to $-5.7\times10^{-19}$ with $0.2\times10^{-19}$ uncertainty through an expanded-beam-waist optical lattice implementation. Comprehensive lattice light shift calibration reaches $6.3\times10^{-19}$ uncertainty via strategic self-comparison measurements that determine magic frequency, $E1$ polarizability, hyperpolarizability, and $E2\text{-}M1$ polarizability. These advancements integrate with a characterized second-order Zeeman coefficient for magnetically insensitive transitions to collectively achieve the reported uncertainty level.

This performance meets the $2\times10^{-18}$ single-clock accuracy requirement for redefining the SI second, with potential applications in relativistic geodesy and high-resolution dark matter searches. Future optical clocks targeting uncertainties in the low-$10^{-19}$ range may benefit from combining cryogenic environments \cite{Hassan2025} with vertical shallow lattice geometries \cite{Kim2023}, which promise further suppression of BBR and lattice light shift uncertainties.

\begin{acknowledgments}

This work is supported by
the Scientific Research Innovation Capability Support Project for Young Faculty (Gtant No.~{ZYGXQNJSKYCXNLZCXM-I26}), 
the National Key Research and Development Program of China (Grants No.2020YFA0309804, SQ2024YFC220046, 2023YFC2206200), 
National Natural Science Foundation of China (Gant No.12204464),
Shanghai Rising-Star Program (Grant No.22QA1409800), 
Independent Deployment Project of HFNL (Grant No. ZB2025010100), 
USTC Research Funds of the Double First-Class Initiative (Grant No.YD9990002023), 
the Fundamental Research Funds for the Central Universities (Grant No. WK9990000156), 
Anhui Initiative in Quantum Information Technologies, 
Shanghai Municipal Science and Technology Major Project (Grant No.2019SHZDZX01), 
the Strategic Priority Research Program of Chinese Academy of Sciences (Grants No. XDB35000000, XDA0520101), 
Innovation Program for Quantum Science and Technology (Grant No.2021ZD0300106 and 2021ZD0302002), 
and the New Corner Stone Science Foundation.
\end{acknowledgments}
{\it $^{\#}$ Z.-P. J., J. L., D.-Q. K. and X. Z. contributed equally to this work.}

\bibliography{ref.bib}

\begin{thebibliography}{60}%
\makeatletter
\providecommand \@ifxundefined [1]{%
 \@ifx{#1\undefined}
}%
\providecommand \@ifnum [1]{%
 \ifnum #1\expandafter \@firstoftwo
 \else \expandafter \@secondoftwo
 \fi
}%
\providecommand \@ifx [1]{%
 \ifx #1\expandafter \@firstoftwo
 \else \expandafter \@secondoftwo
 \fi
}%
\providecommand \natexlab [1]{#1}%
\providecommand \enquote  [1]{``#1''}%
\providecommand \bibnamefont  [1]{#1}%
\providecommand \bibfnamefont [1]{#1}%
\providecommand \citenamefont [1]{#1}%
\providecommand \href@noop [0]{\@secondoftwo}%
\providecommand \href [0]{\begingroup \@sanitize@url \@href}%
\providecommand \@href[1]{\@@startlink{#1}\@@href}%
\providecommand \@@href[1]{\endgroup#1\@@endlink}%
\providecommand \@sanitize@url [0]{\catcode `\\12\catcode `\$12\catcode
  `\&12\catcode `\#12\catcode `\^12\catcode `\_12\catcode `\%12\relax}%
\providecommand \@@startlink[1]{}%
\providecommand \@@endlink[0]{}%
\providecommand \url  [0]{\begingroup\@sanitize@url \@url }%
\providecommand \@url [1]{\endgroup\@href {#1}{\urlprefix }}%
\providecommand \urlprefix  [0]{URL }%
\providecommand \Eprint [0]{\href }%
\providecommand \doibase [0]{https://doi.org/}%
\providecommand \selectlanguage [0]{\@gobble}%
\providecommand \bibinfo  [0]{\@secondoftwo}%
\providecommand \bibfield  [0]{\@secondoftwo}%
\providecommand \translation [1]{[#1]}%
\providecommand \BibitemOpen [0]{}%
\providecommand \bibitemStop [0]{}%
\providecommand \bibitemNoStop [0]{.\EOS\space}%
\providecommand \EOS [0]{\spacefactor3000\relax}%
\providecommand \BibitemShut  [1]{\csname bibitem#1\endcsname}%
\let\auto@bib@innerbib\@empty
\bibitem [{\citenamefont {Ludlow}\ \emph {et~al.}(2015)\citenamefont {Ludlow},
  \citenamefont {Boyd}, \citenamefont {Ye}, \citenamefont {Peik},\ and\
  \citenamefont {Schmidt}}]{Ludlow2015}%
  \BibitemOpen
  \bibfield  {author} {\bibinfo {author} {\bibfnamefont {A.~D.}\ \bibnamefont
  {Ludlow}}, \bibinfo {author} {\bibfnamefont {M.~M.}\ \bibnamefont {Boyd}},
  \bibinfo {author} {\bibfnamefont {J.}~\bibnamefont {Ye}}, \bibinfo {author}
  {\bibfnamefont {E.}~\bibnamefont {Peik}},\ and\ \bibinfo {author}
  {\bibfnamefont {P.~O.}\ \bibnamefont {Schmidt}},\ }\bibfield  {title}
  {\bibinfo {title} {Optical atomic clocks},\ }\href
  {https://doi.org/10.1103/revmodphys.87.637} {\bibfield  {journal} {\bibinfo
  {journal} {Reviews of Modern Physics}\ }\textbf {\bibinfo {volume} {87}},\
  \bibinfo {pages} {637–701} (\bibinfo {year} {2015})}\BibitemShut {NoStop}%
\bibitem [{\citenamefont {McGrew}\ \emph {et~al.}(2018)\citenamefont {McGrew},
  \citenamefont {Zhang}, \citenamefont {Fasano}, \citenamefont {Sch\"{a}ffer},
  \citenamefont {Beloy}, \citenamefont {Nicolodi}, \citenamefont {Brown},
  \citenamefont {Hinkley}, \citenamefont {Milani}, \citenamefont {Schioppo},
  \citenamefont {Yoon},\ and\ \citenamefont {Ludlow}}]{McGrew2018}%
  \BibitemOpen
  \bibfield  {author} {\bibinfo {author} {\bibfnamefont {W.~F.}\ \bibnamefont
  {McGrew}}, \bibinfo {author} {\bibfnamefont {X.}~\bibnamefont {Zhang}},
  \bibinfo {author} {\bibfnamefont {R.~J.}\ \bibnamefont {Fasano}}, \bibinfo
  {author} {\bibfnamefont {S.~A.}\ \bibnamefont {Sch\"{a}ffer}}, \bibinfo
  {author} {\bibfnamefont {K.}~\bibnamefont {Beloy}}, \bibinfo {author}
  {\bibfnamefont {D.}~\bibnamefont {Nicolodi}}, \bibinfo {author}
  {\bibfnamefont {R.~C.}\ \bibnamefont {Brown}}, \bibinfo {author}
  {\bibfnamefont {N.}~\bibnamefont {Hinkley}}, \bibinfo {author} {\bibfnamefont
  {G.}~\bibnamefont {Milani}}, \bibinfo {author} {\bibfnamefont
  {M.}~\bibnamefont {Schioppo}}, \bibinfo {author} {\bibfnamefont {T.~H.}\
  \bibnamefont {Yoon}},\ and\ \bibinfo {author} {\bibfnamefont {A.~D.}\
  \bibnamefont {Ludlow}},\ }\bibfield  {title} {\bibinfo {title} {Atomic clock
  performance enabling geodesy below the centimetre level},\ }\href
  {https://doi.org/10.1038/s41586-018-0738-2} {\bibfield  {journal} {\bibinfo
  {journal} {Nature}\ }\textbf {\bibinfo {volume} {564}},\ \bibinfo {pages}
  {87} (\bibinfo {year} {2018})}\BibitemShut {NoStop}%
\bibitem [{\citenamefont {Takamoto}\ \emph {et~al.}(2020)\citenamefont
  {Takamoto}, \citenamefont {Ushijima}, \citenamefont {Ohmae}, \citenamefont
  {Yahagi}, \citenamefont {Kokado}, \citenamefont {Shinkai},\ and\
  \citenamefont {Katori}}]{Takamoto2020}%
  \BibitemOpen
  \bibfield  {author} {\bibinfo {author} {\bibfnamefont {M.}~\bibnamefont
  {Takamoto}}, \bibinfo {author} {\bibfnamefont {I.}~\bibnamefont {Ushijima}},
  \bibinfo {author} {\bibfnamefont {N.}~\bibnamefont {Ohmae}}, \bibinfo
  {author} {\bibfnamefont {T.}~\bibnamefont {Yahagi}}, \bibinfo {author}
  {\bibfnamefont {K.}~\bibnamefont {Kokado}}, \bibinfo {author} {\bibfnamefont
  {H.}~\bibnamefont {Shinkai}},\ and\ \bibinfo {author} {\bibfnamefont
  {H.}~\bibnamefont {Katori}},\ }\bibfield  {title} {\bibinfo {title} {Test of
  general relativity by a pair of transportable optical lattice clocks},\
  }\href {https://doi.org/10.1038/s41566-020-0619-8} {\bibfield  {journal}
  {\bibinfo  {journal} {Nature Photonics}\ }\textbf {\bibinfo {volume} {14}},\
  \bibinfo {pages} {411} (\bibinfo {year} {2020})}\BibitemShut {NoStop}%
\bibitem [{\citenamefont {Aeppli}\ \emph {et~al.}(2024)\citenamefont {Aeppli},
  \citenamefont {Kim}, \citenamefont {Warfield}, \citenamefont {Safronova},\
  and\ \citenamefont {Ye}}]{Aeppli2024}%
  \BibitemOpen
  \bibfield  {author} {\bibinfo {author} {\bibfnamefont {A.}~\bibnamefont
  {Aeppli}}, \bibinfo {author} {\bibfnamefont {K.}~\bibnamefont {Kim}},
  \bibinfo {author} {\bibfnamefont {W.}~\bibnamefont {Warfield}}, \bibinfo
  {author} {\bibfnamefont {M.~S.}\ \bibnamefont {Safronova}},\ and\ \bibinfo
  {author} {\bibfnamefont {J.}~\bibnamefont {Ye}},\ }\bibfield  {title}
  {\bibinfo {title} {Clock with $8\times10^{-19}$ systematic uncertainty},\
  }\bibfield  {journal} {\bibinfo  {journal} {Physical Review Letters}\
  }\textbf {\bibinfo {volume} {133}},\ \href
  {https://doi.org/10.1103/physrevlett.133.023401}
  {10.1103/physrevlett.133.023401} (\bibinfo {year} {2024})\BibitemShut
  {NoStop}%
\bibitem [{\citenamefont {Marshall}\ \emph {et~al.}(2025)\citenamefont
  {Marshall}, \citenamefont {Castillo}, \citenamefont {Arthur-Dworschack},
  \citenamefont {Aeppli}, \citenamefont {Kim}, \citenamefont {Lee},
  \citenamefont {Warfield}, \citenamefont {Hinrichs}, \citenamefont {Nardelli},
  \citenamefont {Fortier}, \citenamefont {Ye}, \citenamefont {Leibrandt},\ and\
  \citenamefont {Hume}}]{Marshall2025}%
  \BibitemOpen
  \bibfield  {author} {\bibinfo {author} {\bibfnamefont {M.~C.}\ \bibnamefont
  {Marshall}}, \bibinfo {author} {\bibfnamefont {D.~A.~R.}\ \bibnamefont
  {Castillo}}, \bibinfo {author} {\bibfnamefont {W.~J.}\ \bibnamefont
  {Arthur-Dworschack}}, \bibinfo {author} {\bibfnamefont {A.}~\bibnamefont
  {Aeppli}}, \bibinfo {author} {\bibfnamefont {K.}~\bibnamefont {Kim}},
  \bibinfo {author} {\bibfnamefont {D.}~\bibnamefont {Lee}}, \bibinfo {author}
  {\bibfnamefont {W.}~\bibnamefont {Warfield}}, \bibinfo {author}
  {\bibfnamefont {J.}~\bibnamefont {Hinrichs}}, \bibinfo {author}
  {\bibfnamefont {N.~V.}\ \bibnamefont {Nardelli}}, \bibinfo {author}
  {\bibfnamefont {T.~M.}\ \bibnamefont {Fortier}}, \bibinfo {author}
  {\bibfnamefont {J.}~\bibnamefont {Ye}}, \bibinfo {author} {\bibfnamefont
  {D.~R.}\ \bibnamefont {Leibrandt}},\ and\ \bibinfo {author} {\bibfnamefont
  {D.~B.}\ \bibnamefont {Hume}},\ }\bibfield  {title} {\bibinfo {title}
  {High-stability single-ion clock with $5.5\times10^{-19}$ systematic
  uncertainty},\ }\bibfield  {journal} {\bibinfo  {journal} {Physical Review
  Letters}\ }\textbf {\bibinfo {volume} {135}},\ \href
  {https://doi.org/10.1103/hb3c-dk28} {10.1103/hb3c-dk28} (\bibinfo {year}
  {2025})\BibitemShut {NoStop}%
\bibitem [{\citenamefont {Zhang}\ \emph {et~al.}(2023)\citenamefont {Zhang},
  \citenamefont {Arnold}, \citenamefont {Kaewuam},\ and\ \citenamefont
  {Barrett}}]{Zhang2023}%
  \BibitemOpen
  \bibfield  {author} {\bibinfo {author} {\bibfnamefont {Z.}~\bibnamefont
  {Zhang}}, \bibinfo {author} {\bibfnamefont {K.~J.}\ \bibnamefont {Arnold}},
  \bibinfo {author} {\bibfnamefont {R.}~\bibnamefont {Kaewuam}},\ and\ \bibinfo
  {author} {\bibfnamefont {M.~D.}\ \bibnamefont {Barrett}},\ }\bibfield
  {title} {\bibinfo {title} {${}^{176}\text{Lu}^+$ clock comparison at the
  $10^{-18}$ level via correlation spectroscopy},\ }\href
  {https://doi.org/10.1126/sciadv.adg1971} {\bibfield  {journal} {\bibinfo
  {journal} {Science Advances}\ }\textbf {\bibinfo {volume} {9}},\ \bibinfo
  {pages} {eadg1971} (\bibinfo {year} {2023})}\BibitemShut {NoStop}%
\bibitem [{\citenamefont {Tofful}\ \emph {et~al.}(2024)\citenamefont {Tofful},
  \citenamefont {Baynham}, \citenamefont {Curtis}, \citenamefont {Parsons},
  \citenamefont {Robertson}, \citenamefont {Schioppo}, \citenamefont {Tunesi},
  \citenamefont {Margolis}, \citenamefont {Hendricks}, \citenamefont {Whale},
  \citenamefont {Thompson},\ and\ \citenamefont {Godun}}]{Tofful2024}%
  \BibitemOpen
  \bibfield  {author} {\bibinfo {author} {\bibfnamefont {A.}~\bibnamefont
  {Tofful}}, \bibinfo {author} {\bibfnamefont {C.~F.~A.}\ \bibnamefont
  {Baynham}}, \bibinfo {author} {\bibfnamefont {E.~A.}\ \bibnamefont {Curtis}},
  \bibinfo {author} {\bibfnamefont {A.~O.}\ \bibnamefont {Parsons}}, \bibinfo
  {author} {\bibfnamefont {B.~I.}\ \bibnamefont {Robertson}}, \bibinfo {author}
  {\bibfnamefont {M.}~\bibnamefont {Schioppo}}, \bibinfo {author}
  {\bibfnamefont {J.}~\bibnamefont {Tunesi}}, \bibinfo {author} {\bibfnamefont
  {H.~S.}\ \bibnamefont {Margolis}}, \bibinfo {author} {\bibfnamefont {R.~J.}\
  \bibnamefont {Hendricks}}, \bibinfo {author} {\bibfnamefont {J.}~\bibnamefont
  {Whale}}, \bibinfo {author} {\bibfnamefont {R.~C.}\ \bibnamefont
  {Thompson}},\ and\ \bibinfo {author} {\bibfnamefont {R.~M.}\ \bibnamefont
  {Godun}},\ }\bibfield  {title} {\bibinfo {title} {${}^{171}\text{Yb}^+$
  optical clock with $2.2\times 10^{-18}$ systematic uncertainty and absolute
  frequency measurements},\ }\href {https://doi.org/10.1088/1681-7575/ad53cd}
  {\bibfield  {journal} {\bibinfo  {journal} {Metrologia}\ }\textbf {\bibinfo
  {volume} {61}},\ \bibinfo {pages} {045001} (\bibinfo {year}
  {2024})}\BibitemShut {NoStop}%
\bibitem [{\citenamefont {Hausser}\ \emph {et~al.}(2025)\citenamefont
  {Hausser}, \citenamefont {Keller}, \citenamefont {Nordmann}, \citenamefont
  {Bhatt}, \citenamefont {Kiethe}, \citenamefont {Liu}, \citenamefont
  {Richter}, \citenamefont {von Boehn}, \citenamefont {Rahm}, \citenamefont
  {Weyers}, \citenamefont {Benkler}, \citenamefont {Lipphardt}, \citenamefont
  {D\"orscher}, \citenamefont {Stahl}, \citenamefont {Klose}, \citenamefont
  {Lisdat}, \citenamefont {Filzinger}, \citenamefont {Huntemann}, \citenamefont
  {Peik},\ and\ \citenamefont {Mehlst\"aubler}}]{Hausser2025}%
  \BibitemOpen
  \bibfield  {author} {\bibinfo {author} {\bibfnamefont {H.~N.}\ \bibnamefont
  {Hausser}}, \bibinfo {author} {\bibfnamefont {J.}~\bibnamefont {Keller}},
  \bibinfo {author} {\bibfnamefont {T.}~\bibnamefont {Nordmann}}, \bibinfo
  {author} {\bibfnamefont {N.~M.}\ \bibnamefont {Bhatt}}, \bibinfo {author}
  {\bibfnamefont {J.}~\bibnamefont {Kiethe}}, \bibinfo {author} {\bibfnamefont
  {H.}~\bibnamefont {Liu}}, \bibinfo {author} {\bibfnamefont {I.~M.}\
  \bibnamefont {Richter}}, \bibinfo {author} {\bibfnamefont {M.}~\bibnamefont
  {von Boehn}}, \bibinfo {author} {\bibfnamefont {J.}~\bibnamefont {Rahm}},
  \bibinfo {author} {\bibfnamefont {S.}~\bibnamefont {Weyers}}, \bibinfo
  {author} {\bibfnamefont {E.}~\bibnamefont {Benkler}}, \bibinfo {author}
  {\bibfnamefont {B.}~\bibnamefont {Lipphardt}}, \bibinfo {author}
  {\bibfnamefont {S.}~\bibnamefont {D\"orscher}}, \bibinfo {author}
  {\bibfnamefont {K.}~\bibnamefont {Stahl}}, \bibinfo {author} {\bibfnamefont
  {J.}~\bibnamefont {Klose}}, \bibinfo {author} {\bibfnamefont
  {C.}~\bibnamefont {Lisdat}}, \bibinfo {author} {\bibfnamefont
  {M.}~\bibnamefont {Filzinger}}, \bibinfo {author} {\bibfnamefont
  {N.}~\bibnamefont {Huntemann}}, \bibinfo {author} {\bibfnamefont
  {E.}~\bibnamefont {Peik}},\ and\ \bibinfo {author} {\bibfnamefont {T.~E.}\
  \bibnamefont {Mehlst\"aubler}},\ }\bibfield  {title} {\bibinfo {title}
  {{${}^{115}{\mathrm{In}}^{+}\text{-}{}^{172}{\mathrm{Yb}}^{+}$} coulomb
  crystal clock with $2.5\times10^{-18}$ systematic uncertainty},\ }\href
  {https://doi.org/10.1103/PhysRevLett.134.023201} {\bibfield  {journal}
  {\bibinfo  {journal} {Phys. Rev. Lett.}\ }\textbf {\bibinfo {volume} {134}},\
  \bibinfo {pages} {023201} (\bibinfo {year} {2025})}\BibitemShut {NoStop}%
\bibitem [{\citenamefont {Zhang}\ \emph {et~al.}(2025)\citenamefont {Zhang},
  \citenamefont {Ma}, \citenamefont {Huang}, \citenamefont {Han}, \citenamefont
  {Hu}, \citenamefont {Wang}, \citenamefont {Zhang}, \citenamefont {Tang},
  \citenamefont {Shi}, \citenamefont {Guan},\ and\ \citenamefont
  {Gao}}]{Zhang2025}%
  \BibitemOpen
  \bibfield  {author} {\bibinfo {author} {\bibfnamefont {B.}~\bibnamefont
  {Zhang}}, \bibinfo {author} {\bibfnamefont {Z.}~\bibnamefont {Ma}}, \bibinfo
  {author} {\bibfnamefont {Y.}~\bibnamefont {Huang}}, \bibinfo {author}
  {\bibfnamefont {H.}~\bibnamefont {Han}}, \bibinfo {author} {\bibfnamefont
  {R.}~\bibnamefont {Hu}}, \bibinfo {author} {\bibfnamefont {Y.}~\bibnamefont
  {Wang}}, \bibinfo {author} {\bibfnamefont {H.}~\bibnamefont {Zhang}},
  \bibinfo {author} {\bibfnamefont {L.}~\bibnamefont {Tang}}, \bibinfo {author}
  {\bibfnamefont {T.}~\bibnamefont {Shi}}, \bibinfo {author} {\bibfnamefont
  {H.}~\bibnamefont {Guan}},\ and\ \bibinfo {author} {\bibfnamefont
  {K.}~\bibnamefont {Gao}},\ }\href {https://arxiv.org/abs/2506.17423}
  {\bibinfo {title} {A liquid-nitrogen-cooled {${}^{40}\text{Ca}^+$} ion
  optical clock with a systematic uncertainty of $4.6\times 10^{-19}$}}
  (\bibinfo {year} {2025}),\ \Eprint {https://arxiv.org/abs/2506.17423}
  {arXiv:2506.17423 [physics.atom-ph]} \BibitemShut {NoStop}%
\bibitem [{\citenamefont {Dimarcq}\ \emph {et~al.}(2024)\citenamefont
  {Dimarcq}, \citenamefont {Gertsvolf}, \citenamefont {Mileti}, \citenamefont
  {Bize}, \citenamefont {Oates}, \citenamefont {Peik}, \citenamefont
  {Calonico}, \citenamefont {Ido}, \citenamefont {Tavella}, \citenamefont
  {Meynadier}, \citenamefont {Petit}, \citenamefont {Panfilo}, \citenamefont
  {Bartholomew}, \citenamefont {Defraigne}, \citenamefont {Donley},
  \citenamefont {Hedekvist}, \citenamefont {Sesia}, \citenamefont {Wouters},
  \citenamefont {Dubé}, \citenamefont {Fang}, \citenamefont {Levi},
  \citenamefont {Lodewyck}, \citenamefont {Margolis}, \citenamefont {Newell},
  \citenamefont {Slyusarev}, \citenamefont {Weyers}, \citenamefont {Uzan},
  \citenamefont {Yasuda}, \citenamefont {Yu}, \citenamefont {Rieck},
  \citenamefont {Schnatz}, \citenamefont {Hanado}, \citenamefont {Fujieda},
  \citenamefont {Pottie}, \citenamefont {Hanssen}, \citenamefont {Malimon},\
  and\ \citenamefont {Ashby}}]{Dimarcq2024}%
  \BibitemOpen
  \bibfield  {author} {\bibinfo {author} {\bibfnamefont {N.}~\bibnamefont
  {Dimarcq}}, \bibinfo {author} {\bibfnamefont {M.}~\bibnamefont {Gertsvolf}},
  \bibinfo {author} {\bibfnamefont {G.}~\bibnamefont {Mileti}}, \bibinfo
  {author} {\bibfnamefont {S.}~\bibnamefont {Bize}}, \bibinfo {author}
  {\bibfnamefont {C.~W.}\ \bibnamefont {Oates}}, \bibinfo {author}
  {\bibfnamefont {E.}~\bibnamefont {Peik}}, \bibinfo {author} {\bibfnamefont
  {D.}~\bibnamefont {Calonico}}, \bibinfo {author} {\bibfnamefont
  {T.}~\bibnamefont {Ido}}, \bibinfo {author} {\bibfnamefont {P.}~\bibnamefont
  {Tavella}}, \bibinfo {author} {\bibfnamefont {F.}~\bibnamefont {Meynadier}},
  \bibinfo {author} {\bibfnamefont {G.}~\bibnamefont {Petit}}, \bibinfo
  {author} {\bibfnamefont {G.}~\bibnamefont {Panfilo}}, \bibinfo {author}
  {\bibfnamefont {J.}~\bibnamefont {Bartholomew}}, \bibinfo {author}
  {\bibfnamefont {P.}~\bibnamefont {Defraigne}}, \bibinfo {author}
  {\bibfnamefont {E.~A.}\ \bibnamefont {Donley}}, \bibinfo {author}
  {\bibfnamefont {P.~O.}\ \bibnamefont {Hedekvist}}, \bibinfo {author}
  {\bibfnamefont {I.}~\bibnamefont {Sesia}}, \bibinfo {author} {\bibfnamefont
  {M.}~\bibnamefont {Wouters}}, \bibinfo {author} {\bibfnamefont
  {P.}~\bibnamefont {Dubé}}, \bibinfo {author} {\bibfnamefont
  {F.}~\bibnamefont {Fang}}, \bibinfo {author} {\bibfnamefont {F.}~\bibnamefont
  {Levi}}, \bibinfo {author} {\bibfnamefont {J.}~\bibnamefont {Lodewyck}},
  \bibinfo {author} {\bibfnamefont {H.~S.}\ \bibnamefont {Margolis}}, \bibinfo
  {author} {\bibfnamefont {D.}~\bibnamefont {Newell}}, \bibinfo {author}
  {\bibfnamefont {S.}~\bibnamefont {Slyusarev}}, \bibinfo {author}
  {\bibfnamefont {S.}~\bibnamefont {Weyers}}, \bibinfo {author} {\bibfnamefont
  {J.-P.}\ \bibnamefont {Uzan}}, \bibinfo {author} {\bibfnamefont
  {M.}~\bibnamefont {Yasuda}}, \bibinfo {author} {\bibfnamefont {D.-H.}\
  \bibnamefont {Yu}}, \bibinfo {author} {\bibfnamefont {C.}~\bibnamefont
  {Rieck}}, \bibinfo {author} {\bibfnamefont {H.}~\bibnamefont {Schnatz}},
  \bibinfo {author} {\bibfnamefont {Y.}~\bibnamefont {Hanado}}, \bibinfo
  {author} {\bibfnamefont {M.}~\bibnamefont {Fujieda}}, \bibinfo {author}
  {\bibfnamefont {P.-E.}\ \bibnamefont {Pottie}}, \bibinfo {author}
  {\bibfnamefont {J.}~\bibnamefont {Hanssen}}, \bibinfo {author} {\bibfnamefont
  {A.}~\bibnamefont {Malimon}},\ and\ \bibinfo {author} {\bibfnamefont
  {N.}~\bibnamefont {Ashby}},\ }\bibfield  {title} {\bibinfo {title} {Roadmap
  towards the redefinition of the second},\ }\href
  {https://doi.org/10.1088/1681-7575/ad17d2} {\bibfield  {journal} {\bibinfo
  {journal} {Metrologia}\ }\textbf {\bibinfo {volume} {61}},\ \bibinfo {pages}
  {012001} (\bibinfo {year} {2024})}\BibitemShut {NoStop}%
\bibitem [{\citenamefont {Lu}\ \emph {et~al.}(2025)\citenamefont {Lu},
  \citenamefont {Guo}, \citenamefont {Liu}, \citenamefont {Cao}, \citenamefont
  {Li}, \citenamefont {Xia}, \citenamefont {Xu}, \citenamefont {Lu},
  \citenamefont {Wang},\ and\ \citenamefont {Chang}}]{Lu2025}%
  \BibitemOpen
  \bibfield  {author} {\bibinfo {author} {\bibfnamefont {X.-T.}\ \bibnamefont
  {Lu}}, \bibinfo {author} {\bibfnamefont {F.}~\bibnamefont {Guo}}, \bibinfo
  {author} {\bibfnamefont {Y.-Y.}\ \bibnamefont {Liu}}, \bibinfo {author}
  {\bibfnamefont {J.}~\bibnamefont {Cao}}, \bibinfo {author} {\bibfnamefont
  {J.-A.}\ \bibnamefont {Li}}, \bibinfo {author} {\bibfnamefont {J.-J.}\
  \bibnamefont {Xia}}, \bibinfo {author} {\bibfnamefont {Q.-F.}\ \bibnamefont
  {Xu}}, \bibinfo {author} {\bibfnamefont {B.-Q.}\ \bibnamefont {Lu}}, \bibinfo
  {author} {\bibfnamefont {Y.-B.}\ \bibnamefont {Wang}},\ and\ \bibinfo
  {author} {\bibfnamefont {H.}~\bibnamefont {Chang}},\ }\bibfield  {title}
  {\bibinfo {title} {{NTSC SrII} optical lattice clock with uncertainty of
  $2\times10^{-18}$},\ }\href {https://doi.org/10.1088/1681-7575/addc77}
  {\bibfield  {journal} {\bibinfo  {journal} {Metrologia}\ }\textbf {\bibinfo
  {volume} {62}},\ \bibinfo {pages} {035007} (\bibinfo {year}
  {2025})}\BibitemShut {NoStop}%
\bibitem [{\citenamefont {Ma}\ \emph {et~al.}(2024)\citenamefont {Ma},
  \citenamefont {Deng}, \citenamefont {Wang}, \citenamefont {Wei},
  \citenamefont {Hao}, \citenamefont {Zhang}, \citenamefont {Pang},
  \citenamefont {Wang}, \citenamefont {Wu}, \citenamefont {Liu}, \citenamefont
  {Yuan}, \citenamefont {Chang}, \citenamefont {Zhang}, \citenamefont {Wu},
  \citenamefont {Zhang},\ and\ \citenamefont {Lu}}]{Ma2024}%
  \BibitemOpen
  \bibfield  {author} {\bibinfo {author} {\bibfnamefont {Z.~Y.}\ \bibnamefont
  {Ma}}, \bibinfo {author} {\bibfnamefont {K.}~\bibnamefont {Deng}}, \bibinfo
  {author} {\bibfnamefont {Z.~Y.}\ \bibnamefont {Wang}}, \bibinfo {author}
  {\bibfnamefont {W.~Z.}\ \bibnamefont {Wei}}, \bibinfo {author} {\bibfnamefont
  {P.}~\bibnamefont {Hao}}, \bibinfo {author} {\bibfnamefont {H.~X.}\
  \bibnamefont {Zhang}}, \bibinfo {author} {\bibfnamefont {L.~R.}\ \bibnamefont
  {Pang}}, \bibinfo {author} {\bibfnamefont {B.}~\bibnamefont {Wang}}, \bibinfo
  {author} {\bibfnamefont {F.~F.}\ \bibnamefont {Wu}}, \bibinfo {author}
  {\bibfnamefont {H.~L.}\ \bibnamefont {Liu}}, \bibinfo {author} {\bibfnamefont
  {W.~H.}\ \bibnamefont {Yuan}}, \bibinfo {author} {\bibfnamefont {J.~L.}\
  \bibnamefont {Chang}}, \bibinfo {author} {\bibfnamefont {J.~X.}\ \bibnamefont
  {Zhang}}, \bibinfo {author} {\bibfnamefont {Q.~Y.}\ \bibnamefont {Wu}},
  \bibinfo {author} {\bibfnamefont {J.}~\bibnamefont {Zhang}},\ and\ \bibinfo
  {author} {\bibfnamefont {Z.~H.}\ \bibnamefont {Lu}},\ }\bibfield  {title}
  {\bibinfo {title} {Quantum-logic-based
  ${}^{25}\text{Mg}^+$-${}^{27}\text{Al}^+$ optical frequency standard for the
  redefinition of the si second},\ }\bibfield  {journal} {\bibinfo  {journal}
  {Physical Review Applied}\ }\textbf {\bibinfo {volume} {21}},\ \href
  {https://doi.org/10.1103/physrevapplied.21.044017}
  {10.1103/physrevapplied.21.044017} (\bibinfo {year} {2024})\BibitemShut
  {NoStop}%
\bibitem [{\citenamefont {Ludlow}\ \emph {et~al.}(2006)\citenamefont {Ludlow},
  \citenamefont {Boyd}, \citenamefont {Zelevinsky}, \citenamefont {Foreman},
  \citenamefont {Blatt}, \citenamefont {Notcutt}, \citenamefont {Ido},\ and\
  \citenamefont {Ye}}]{Ludlow2006}%
  \BibitemOpen
  \bibfield  {author} {\bibinfo {author} {\bibfnamefont {A.~D.}\ \bibnamefont
  {Ludlow}}, \bibinfo {author} {\bibfnamefont {M.~M.}\ \bibnamefont {Boyd}},
  \bibinfo {author} {\bibfnamefont {T.}~\bibnamefont {Zelevinsky}}, \bibinfo
  {author} {\bibfnamefont {S.~M.}\ \bibnamefont {Foreman}}, \bibinfo {author}
  {\bibfnamefont {S.}~\bibnamefont {Blatt}}, \bibinfo {author} {\bibfnamefont
  {M.}~\bibnamefont {Notcutt}}, \bibinfo {author} {\bibfnamefont
  {T.}~\bibnamefont {Ido}},\ and\ \bibinfo {author} {\bibfnamefont
  {J.}~\bibnamefont {Ye}},\ }\bibfield  {title} {\bibinfo {title} {Systematic
  study of the ${}^{87}${Sr} clock transition in an optical lattice},\
  }\bibfield  {journal} {\bibinfo  {journal} {Physical Review Letters}\
  }\textbf {\bibinfo {volume} {96}},\ \href
  {https://doi.org/10.1103/physrevlett.96.033003}
  {10.1103/physrevlett.96.033003} (\bibinfo {year} {2006})\BibitemShut
  {NoStop}%
\bibitem [{\citenamefont {Bothwell}\ \emph {et~al.}(2019)\citenamefont
  {Bothwell}, \citenamefont {Kedar}, \citenamefont {Oelker}, \citenamefont
  {Robinson}, \citenamefont {Bromley}, \citenamefont {Tew}, \citenamefont
  {Ye},\ and\ \citenamefont {Kennedy}}]{Bothwell2019}%
  \BibitemOpen
  \bibfield  {author} {\bibinfo {author} {\bibfnamefont {T.}~\bibnamefont
  {Bothwell}}, \bibinfo {author} {\bibfnamefont {D.}~\bibnamefont {Kedar}},
  \bibinfo {author} {\bibfnamefont {E.}~\bibnamefont {Oelker}}, \bibinfo
  {author} {\bibfnamefont {J.~M.}\ \bibnamefont {Robinson}}, \bibinfo {author}
  {\bibfnamefont {S.~L.}\ \bibnamefont {Bromley}}, \bibinfo {author}
  {\bibfnamefont {W.~L.}\ \bibnamefont {Tew}}, \bibinfo {author} {\bibfnamefont
  {J.}~\bibnamefont {Ye}},\ and\ \bibinfo {author} {\bibfnamefont {C.~J.}\
  \bibnamefont {Kennedy}},\ }\bibfield  {title} {\bibinfo {title} {{JILA} {SrI}
  optical lattice clock with uncertainty of $2.0 \times 10^{-18}$},\ }\href
  {https://doi.org/10.1088/1681-7575/ab4089} {\bibfield  {journal} {\bibinfo
  {journal} {Metrologia}\ }\textbf {\bibinfo {volume} {56}},\ \bibinfo {pages}
  {065004} (\bibinfo {year} {2019})}\BibitemShut {NoStop}%
\bibitem [{\citenamefont {Westergaard}\ \emph {et~al.}(2011)\citenamefont
  {Westergaard}, \citenamefont {Lodewyck}, \citenamefont {Lorini},
  \citenamefont {Lecallier}, \citenamefont {Burt}, \citenamefont {Zawada},
  \citenamefont {Millo},\ and\ \citenamefont {Lemonde}}]{Westergaard2011}%
  \BibitemOpen
  \bibfield  {author} {\bibinfo {author} {\bibfnamefont {P.~G.}\ \bibnamefont
  {Westergaard}}, \bibinfo {author} {\bibfnamefont {J.}~\bibnamefont
  {Lodewyck}}, \bibinfo {author} {\bibfnamefont {L.}~\bibnamefont {Lorini}},
  \bibinfo {author} {\bibfnamefont {A.}~\bibnamefont {Lecallier}}, \bibinfo
  {author} {\bibfnamefont {E.~A.}\ \bibnamefont {Burt}}, \bibinfo {author}
  {\bibfnamefont {M.}~\bibnamefont {Zawada}}, \bibinfo {author} {\bibfnamefont
  {J.}~\bibnamefont {Millo}},\ and\ \bibinfo {author} {\bibfnamefont
  {P.}~\bibnamefont {Lemonde}},\ }\bibfield  {title} {\bibinfo {title}
  {Lattice-induced frequency shifts in {Sr} optical lattice clocks at the
  $10^{-17}$ level},\ }\bibfield  {journal} {\bibinfo  {journal} {Physical
  Review Letters}\ }\textbf {\bibinfo {volume} {106}},\ \href
  {https://doi.org/10.1103/physrevlett.106.210801}
  {10.1103/physrevlett.106.210801} (\bibinfo {year} {2011})\BibitemShut
  {NoStop}%
\bibitem [{\citenamefont {Safronova}\ \emph {et~al.}(2013)\citenamefont
  {Safronova}, \citenamefont {Porsev}, \citenamefont {Safronova}, \citenamefont
  {Kozlov},\ and\ \citenamefont {Clark}}]{Safronova2013}%
  \BibitemOpen
  \bibfield  {author} {\bibinfo {author} {\bibfnamefont {M.~S.}\ \bibnamefont
  {Safronova}}, \bibinfo {author} {\bibfnamefont {S.~G.}\ \bibnamefont
  {Porsev}}, \bibinfo {author} {\bibfnamefont {U.~I.}\ \bibnamefont
  {Safronova}}, \bibinfo {author} {\bibfnamefont {M.~G.}\ \bibnamefont
  {Kozlov}},\ and\ \bibinfo {author} {\bibfnamefont {C.~W.}\ \bibnamefont
  {Clark}},\ }\bibfield  {title} {\bibinfo {title} {Blackbody-radiation shift
  in the {Sr} optical atomic clock},\ }\bibfield  {journal} {\bibinfo
  {journal} {Physical Review A}\ }\textbf {\bibinfo {volume} {87}},\ \href
  {https://doi.org/10.1103/physreva.87.012509} {10.1103/physreva.87.012509}
  (\bibinfo {year} {2013})\BibitemShut {NoStop}%
\bibitem [{\citenamefont {Brown}\ \emph {et~al.}(2017)\citenamefont {Brown},
  \citenamefont {Phillips}, \citenamefont {Beloy}, \citenamefont {McGrew},
  \citenamefont {Schioppo}, \citenamefont {Fasano}, \citenamefont {Milani},
  \citenamefont {Zhang}, \citenamefont {Hinkley}, \citenamefont {Leopardi},
  \citenamefont {Yoon}, \citenamefont {Nicolodi}, \citenamefont {Fortier},\
  and\ \citenamefont {Ludlow}}]{Brown2017}%
  \BibitemOpen
  \bibfield  {author} {\bibinfo {author} {\bibfnamefont {R.~C.}\ \bibnamefont
  {Brown}}, \bibinfo {author} {\bibfnamefont {N.~B.}\ \bibnamefont {Phillips}},
  \bibinfo {author} {\bibfnamefont {K.}~\bibnamefont {Beloy}}, \bibinfo
  {author} {\bibfnamefont {W.}~\bibnamefont {McGrew}}, \bibinfo {author}
  {\bibfnamefont {M.}~\bibnamefont {Schioppo}}, \bibinfo {author}
  {\bibfnamefont {R.~J.}\ \bibnamefont {Fasano}}, \bibinfo {author}
  {\bibfnamefont {G.}~\bibnamefont {Milani}}, \bibinfo {author} {\bibfnamefont
  {X.}~\bibnamefont {Zhang}}, \bibinfo {author} {\bibfnamefont
  {N.}~\bibnamefont {Hinkley}}, \bibinfo {author} {\bibfnamefont
  {H.}~\bibnamefont {Leopardi}}, \bibinfo {author} {\bibfnamefont {T.~H.}\
  \bibnamefont {Yoon}}, \bibinfo {author} {\bibfnamefont {D.}~\bibnamefont
  {Nicolodi}}, \bibinfo {author} {\bibfnamefont {T.~M.}\ \bibnamefont
  {Fortier}},\ and\ \bibinfo {author} {\bibfnamefont {A.~D.}\ \bibnamefont
  {Ludlow}},\ }\bibfield  {title} {\bibinfo {title} {Hyperpolarizability and
  operational magic wavelength in an optical lattice clock},\ }\bibfield
  {journal} {\bibinfo  {journal} {Phys. Rev. Lett.}\ }\textbf {\bibinfo
  {volume} {119}},\ \href {https://doi.org/10.1103/physrevlett.119.253001}
  {10.1103/physrevlett.119.253001} (\bibinfo {year} {2017})\BibitemShut
  {NoStop}%
\bibitem [{\citenamefont {Ushijima}\ \emph {et~al.}(2018)\citenamefont
  {Ushijima}, \citenamefont {Takamoto},\ and\ \citenamefont
  {Katori}}]{Ushijima2018}%
  \BibitemOpen
  \bibfield  {author} {\bibinfo {author} {\bibfnamefont {I.}~\bibnamefont
  {Ushijima}}, \bibinfo {author} {\bibfnamefont {M.}~\bibnamefont {Takamoto}},\
  and\ \bibinfo {author} {\bibfnamefont {H.}~\bibnamefont {Katori}},\
  }\bibfield  {title} {\bibinfo {title} {Operational magic intensity for {Sr}
  optical lattice clocks},\ }\href
  {https://doi.org/10.1103/PhysRevLett.121.263202} {\bibfield  {journal}
  {\bibinfo  {journal} {Phys. Rev. Lett.}\ }\textbf {\bibinfo {volume} {121}},\
  \bibinfo {pages} {263202} (\bibinfo {year} {2018})}\BibitemShut {NoStop}%
\bibitem [{\citenamefont {Nemitz}\ \emph {et~al.}(2019)\citenamefont {Nemitz},
  \citenamefont {Jørgensen}, \citenamefont {Yanagimoto}, \citenamefont
  {Bregolin},\ and\ \citenamefont {Katori}}]{Nemitz2019}%
  \BibitemOpen
  \bibfield  {author} {\bibinfo {author} {\bibfnamefont {N.}~\bibnamefont
  {Nemitz}}, \bibinfo {author} {\bibfnamefont {A.~A.}\ \bibnamefont
  {Jørgensen}}, \bibinfo {author} {\bibfnamefont {R.}~\bibnamefont
  {Yanagimoto}}, \bibinfo {author} {\bibfnamefont {F.}~\bibnamefont
  {Bregolin}},\ and\ \bibinfo {author} {\bibfnamefont {H.}~\bibnamefont
  {Katori}},\ }\bibfield  {title} {\bibinfo {title} {Modeling light shifts in
  optical lattice clocks},\ }\bibfield  {journal} {\bibinfo  {journal}
  {Physical Review A}\ }\textbf {\bibinfo {volume} {99}},\ \href
  {https://doi.org/10.1103/physreva.99.033424} {10.1103/physreva.99.033424}
  (\bibinfo {year} {2019})\BibitemShut {NoStop}%
\bibitem [{\citenamefont {Lisdat}\ \emph {et~al.}(2021)\citenamefont {Lisdat},
  \citenamefont {Drscher}, \citenamefont {Nosske},\ and\ \citenamefont
  {Sterr}}]{Lisdat2021}%
  \BibitemOpen
  \bibfield  {author} {\bibinfo {author} {\bibfnamefont {C.}~\bibnamefont
  {Lisdat}}, \bibinfo {author} {\bibfnamefont {S.}~\bibnamefont {Drscher}},
  \bibinfo {author} {\bibfnamefont {I.}~\bibnamefont {Nosske}},\ and\ \bibinfo
  {author} {\bibfnamefont {U.}~\bibnamefont {Sterr}},\ }\bibfield  {title}
  {\bibinfo {title} {Blackbody radiation shift in strontium lattice clocks
  revisited},\ }\href {https://doi.org/10.1103/PhysRevResearch.3.L042036}
  {\bibfield  {journal} {\bibinfo  {journal} {Physical Review Research}\
  }\textbf {\bibinfo {volume} {3}},\ \bibinfo {pages} {L042036} (\bibinfo
  {year} {2021})}\BibitemShut {NoStop}%
\bibitem [{\citenamefont {Kim}\ \emph {et~al.}(2023)\citenamefont {Kim},
  \citenamefont {Aeppli}, \citenamefont {Bothwell},\ and\ \citenamefont
  {Ye}}]{Kim2023}%
  \BibitemOpen
  \bibfield  {author} {\bibinfo {author} {\bibfnamefont {K.}~\bibnamefont
  {Kim}}, \bibinfo {author} {\bibfnamefont {A.}~\bibnamefont {Aeppli}},
  \bibinfo {author} {\bibfnamefont {T.}~\bibnamefont {Bothwell}},\ and\
  \bibinfo {author} {\bibfnamefont {J.}~\bibnamefont {Ye}},\ }\bibfield
  {title} {\bibinfo {title} {Evaluation of lattice light shift at low
  $10^{-19}$ uncertainty for a shallow lattice {Sr} optical clock},\ }\bibfield
   {journal} {\bibinfo  {journal} {Physical Review Letters}\ }\textbf {\bibinfo
  {volume} {130}},\ \href {https://doi.org/10.1103/physrevlett.130.113203}
  {10.1103/physrevlett.130.113203} (\bibinfo {year} {2023})\BibitemShut
  {NoStop}%
\bibitem [{\citenamefont {Zhang}\ \emph {et~al.}(2021)\citenamefont {Zhang},
  \citenamefont {Cao}, \citenamefont {Yuan}, \citenamefont {Liu}, \citenamefont
  {Yuan}, \citenamefont {Wei}, \citenamefont {Shu},\ and\ \citenamefont
  {Huang}}]{Zhang2021}%
  \BibitemOpen
  \bibfield  {author} {\bibinfo {author} {\bibfnamefont {P.}~\bibnamefont
  {Zhang}}, \bibinfo {author} {\bibfnamefont {J.}~\bibnamefont {Cao}}, \bibinfo
  {author} {\bibfnamefont {J.-b.}\ \bibnamefont {Yuan}}, \bibinfo {author}
  {\bibfnamefont {D.-x.}\ \bibnamefont {Liu}}, \bibinfo {author} {\bibfnamefont
  {Y.}~\bibnamefont {Yuan}}, \bibinfo {author} {\bibfnamefont {Y.-f.}\
  \bibnamefont {Wei}}, \bibinfo {author} {\bibfnamefont {H.-l.}\ \bibnamefont
  {Shu}},\ and\ \bibinfo {author} {\bibfnamefont {X.-r.}\ \bibnamefont
  {Huang}},\ }\bibfield  {title} {\bibinfo {title} {Evaluation of blackbody
  radiation shift with $2\times10^{-18}$ uncertainty at room temperature for a
  transportable ${}^{40}\text{Ca}^+$ optical clock},\ }\href
  {https://doi.org/10.1088/1681-7575/abe9c4} {\bibfield  {journal} {\bibinfo
  {journal} {Metrologia}\ }\textbf {\bibinfo {volume} {58}},\ \bibinfo {pages}
  {035001} (\bibinfo {year} {2021})}\BibitemShut {NoStop}%
\bibitem [{\citenamefont {Zhang}\ \emph {et~al.}(2022)\citenamefont {Zhang},
  \citenamefont {Xiong}, \citenamefont {Chen}, \citenamefont {Jiang},
  \citenamefont {Wang}, \citenamefont {Tian}, \citenamefont {Zhu},
  \citenamefont {Wang}, \citenamefont {Xiong}, \citenamefont {He},
  \citenamefont {Ma},\ and\ \citenamefont {Lyu}}]{Zhang2022}%
  \BibitemOpen
  \bibfield  {author} {\bibinfo {author} {\bibfnamefont {A.}~\bibnamefont
  {Zhang}}, \bibinfo {author} {\bibfnamefont {Z.}~\bibnamefont {Xiong}},
  \bibinfo {author} {\bibfnamefont {X.}~\bibnamefont {Chen}}, \bibinfo {author}
  {\bibfnamefont {Y.}~\bibnamefont {Jiang}}, \bibinfo {author} {\bibfnamefont
  {J.}~\bibnamefont {Wang}}, \bibinfo {author} {\bibfnamefont {C.}~\bibnamefont
  {Tian}}, \bibinfo {author} {\bibfnamefont {Q.}~\bibnamefont {Zhu}}, \bibinfo
  {author} {\bibfnamefont {B.}~\bibnamefont {Wang}}, \bibinfo {author}
  {\bibfnamefont {D.}~\bibnamefont {Xiong}}, \bibinfo {author} {\bibfnamefont
  {L.}~\bibnamefont {He}}, \bibinfo {author} {\bibfnamefont {L.}~\bibnamefont
  {Ma}},\ and\ \bibinfo {author} {\bibfnamefont {B.}~\bibnamefont {Lyu}},\
  }\bibfield  {title} {\bibinfo {title} {Ytterbium optical lattice clock with
  instability of order $10^{-18}$},\ }\href
  {https://doi.org/10.1088/1681-7575/ac99e4} {\bibfield  {journal} {\bibinfo
  {journal} {Metrologia}\ }\textbf {\bibinfo {volume} {59}},\ \bibinfo {pages}
  {065009} (\bibinfo {year} {2022})}\BibitemShut {NoStop}%
\bibitem [{\citenamefont {Hassan}\ \emph {et~al.}(2025)\citenamefont {Hassan},
  \citenamefont {Beloy}, \citenamefont {Siegel}, \citenamefont {Kobayashi},
  \citenamefont {Swiler}, \citenamefont {Grogan}, \citenamefont {Brown},
  \citenamefont {Rojo}, \citenamefont {Bothwell}, \citenamefont {Hunt},
  \citenamefont {Halaoui},\ and\ \citenamefont {Ludlow}}]{Hassan2025}%
  \BibitemOpen
  \bibfield  {author} {\bibinfo {author} {\bibfnamefont {Y.~S.}\ \bibnamefont
  {Hassan}}, \bibinfo {author} {\bibfnamefont {K.}~\bibnamefont {Beloy}},
  \bibinfo {author} {\bibfnamefont {J.~L.}\ \bibnamefont {Siegel}}, \bibinfo
  {author} {\bibfnamefont {T.}~\bibnamefont {Kobayashi}}, \bibinfo {author}
  {\bibfnamefont {E.}~\bibnamefont {Swiler}}, \bibinfo {author} {\bibfnamefont
  {T.}~\bibnamefont {Grogan}}, \bibinfo {author} {\bibfnamefont {R.~C.}\
  \bibnamefont {Brown}}, \bibinfo {author} {\bibfnamefont {T.}~\bibnamefont
  {Rojo}}, \bibinfo {author} {\bibfnamefont {T.}~\bibnamefont {Bothwell}},
  \bibinfo {author} {\bibfnamefont {B.~D.}\ \bibnamefont {Hunt}}, \bibinfo
  {author} {\bibfnamefont {A.}~\bibnamefont {Halaoui}},\ and\ \bibinfo {author}
  {\bibfnamefont {A.~D.}\ \bibnamefont {Ludlow}},\ }\bibfield  {title}
  {\bibinfo {title} {Cryogenic optical lattice clock with $1.7\times10^{-20}$
  blackbody radiation stark uncertainty},\ }\bibfield  {journal} {\bibinfo
  {journal} {Physical Review Letters}\ }\textbf {\bibinfo {volume} {135}},\
  \href {https://doi.org/10.1103/4tky-jmsm} {10.1103/4tky-jmsm} (\bibinfo
  {year} {2025})\BibitemShut {NoStop}%
\bibitem [{\citenamefont {Jin}\ \emph {et~al.}(2023)\citenamefont {Jin},
  \citenamefont {Zhang}, \citenamefont {Luo}, \citenamefont {Liu},
  \citenamefont {Zhou},\ and\ \citenamefont {Xu}}]{Jin2023}%
  \BibitemOpen
  \bibfield  {author} {\bibinfo {author} {\bibfnamefont {T.}~\bibnamefont
  {Jin}}, \bibinfo {author} {\bibfnamefont {T.}~\bibnamefont {Zhang}}, \bibinfo
  {author} {\bibfnamefont {L.}~\bibnamefont {Luo}}, \bibinfo {author}
  {\bibfnamefont {L.}~\bibnamefont {Liu}}, \bibinfo {author} {\bibfnamefont
  {M.}~\bibnamefont {Zhou}},\ and\ \bibinfo {author} {\bibfnamefont
  {X.}~\bibnamefont {Xu}},\ }\bibfield  {title} {\bibinfo {title} {Multiple
  strategies for evaluating the uncertainty of blackbody radiation frequency
  shift in an optical clock},\ }\href
  {https://doi.org/10.1016/j.measurement.2023.112946} {\bibfield  {journal}
  {\bibinfo  {journal} {Measurement}\ }\textbf {\bibinfo {volume} {216}},\
  \bibinfo {pages} {112946} (\bibinfo {year} {2023})}\BibitemShut {NoStop}%
\bibitem [{\citenamefont {Nicholson}\ \emph {et~al.}(2015)\citenamefont
  {Nicholson}, \citenamefont {Campbell}, \citenamefont {Hutson}, \citenamefont
  {Marti}, \citenamefont {Bloom}, \citenamefont {McNally}, \citenamefont
  {Zhang}, \citenamefont {Barrett}, \citenamefont {Safronova}, \citenamefont
  {Strouse}, \citenamefont {Tew},\ and\ \citenamefont {Ye}}]{Nicholson2015}%
  \BibitemOpen
  \bibfield  {author} {\bibinfo {author} {\bibfnamefont {T.}~\bibnamefont
  {Nicholson}}, \bibinfo {author} {\bibfnamefont {S.}~\bibnamefont {Campbell}},
  \bibinfo {author} {\bibfnamefont {R.}~\bibnamefont {Hutson}}, \bibinfo
  {author} {\bibfnamefont {G.}~\bibnamefont {Marti}}, \bibinfo {author}
  {\bibfnamefont {B.}~\bibnamefont {Bloom}}, \bibinfo {author} {\bibfnamefont
  {R.}~\bibnamefont {McNally}}, \bibinfo {author} {\bibfnamefont
  {W.}~\bibnamefont {Zhang}}, \bibinfo {author} {\bibfnamefont
  {M.}~\bibnamefont {Barrett}}, \bibinfo {author} {\bibfnamefont
  {M.}~\bibnamefont {Safronova}}, \bibinfo {author} {\bibfnamefont
  {G.}~\bibnamefont {Strouse}}, \bibinfo {author} {\bibfnamefont
  {W.}~\bibnamefont {Tew}},\ and\ \bibinfo {author} {\bibfnamefont
  {J.}~\bibnamefont {Ye}},\ }\bibfield  {title} {\bibinfo {title} {Systematic
  evaluation of an atomic clock at $2\times10^{-18}$ total uncertainty},\
  }\href {https://doi.org/10.1038/ncomms7896} {\bibfield  {journal} {\bibinfo
  {journal} {Nature Communications}\ }\textbf {\bibinfo {volume} {6}},\
  \bibinfo {pages} {6896} (\bibinfo {year} {2015})}\BibitemShut {NoStop}%
\bibitem [{\citenamefont {Lu}\ \emph {et~al.}(2022)\citenamefont {Lu},
  \citenamefont {Sun}, \citenamefont {Yang}, \citenamefont {Lin}, \citenamefont
  {Wang}, \citenamefont {Li}, \citenamefont {Meng}, \citenamefont {Lin},
  \citenamefont {Li},\ and\ \citenamefont {Fang}}]{Lu2022}%
  \BibitemOpen
  \bibfield  {author} {\bibinfo {author} {\bibfnamefont {B.-K.}\ \bibnamefont
  {Lu}}, \bibinfo {author} {\bibfnamefont {Z.}~\bibnamefont {Sun}}, \bibinfo
  {author} {\bibfnamefont {T.}~\bibnamefont {Yang}}, \bibinfo {author}
  {\bibfnamefont {Y.-G.}\ \bibnamefont {Lin}}, \bibinfo {author} {\bibfnamefont
  {Q.}~\bibnamefont {Wang}}, \bibinfo {author} {\bibfnamefont {Y.}~\bibnamefont
  {Li}}, \bibinfo {author} {\bibfnamefont {F.}~\bibnamefont {Meng}}, \bibinfo
  {author} {\bibfnamefont {B.-K.}\ \bibnamefont {Lin}}, \bibinfo {author}
  {\bibfnamefont {T.-C.}\ \bibnamefont {Li}},\ and\ \bibinfo {author}
  {\bibfnamefont {Z.-J.}\ \bibnamefont {Fang}},\ }\bibfield  {title} {\bibinfo
  {title} {Improved evaluation of bbr and collisional frequency shifts of
  {NIM-Sr2} with $7.2\times10^{-18}$ total uncertainty},\ }\href
  {https://doi.org/10.1088/0256-307x/39/8/080601} {\bibfield  {journal}
  {\bibinfo  {journal} {Chinese Physics Letters}\ }\textbf {\bibinfo {volume}
  {39}},\ \bibinfo {pages} {080601} (\bibinfo {year} {2022})}\BibitemShut
  {NoStop}%
\bibitem [{\citenamefont {Li}\ \emph {et~al.}(2024)\citenamefont {Li},
  \citenamefont {Cui}, \citenamefont {Jia}, \citenamefont {Kong}, \citenamefont
  {Yu}, \citenamefont {Zhu}, \citenamefont {Liu}, \citenamefont {Wang},
  \citenamefont {Zhang}, \citenamefont {Huang}, \citenamefont {Zhu},
  \citenamefont {Yang}, \citenamefont {Hu}, \citenamefont {Liu}, \citenamefont
  {Zhai}, \citenamefont {Liu}, \citenamefont {Jiang}, \citenamefont {Xu},
  \citenamefont {Dai}, \citenamefont {Chen},\ and\ \citenamefont
  {Pan}}]{Li2024}%
  \BibitemOpen
  \bibfield  {author} {\bibinfo {author} {\bibfnamefont {J.}~\bibnamefont
  {Li}}, \bibinfo {author} {\bibfnamefont {X.-Y.}\ \bibnamefont {Cui}},
  \bibinfo {author} {\bibfnamefont {Z.-P.}\ \bibnamefont {Jia}}, \bibinfo
  {author} {\bibfnamefont {D.-Q.}\ \bibnamefont {Kong}}, \bibinfo {author}
  {\bibfnamefont {H.-W.}\ \bibnamefont {Yu}}, \bibinfo {author} {\bibfnamefont
  {X.-Q.}\ \bibnamefont {Zhu}}, \bibinfo {author} {\bibfnamefont {X.-Y.}\
  \bibnamefont {Liu}}, \bibinfo {author} {\bibfnamefont {D.-Z.}\ \bibnamefont
  {Wang}}, \bibinfo {author} {\bibfnamefont {X.}~\bibnamefont {Zhang}},
  \bibinfo {author} {\bibfnamefont {X.-Y.}\ \bibnamefont {Huang}}, \bibinfo
  {author} {\bibfnamefont {M.-Y.}\ \bibnamefont {Zhu}}, \bibinfo {author}
  {\bibfnamefont {Y.-M.}\ \bibnamefont {Yang}}, \bibinfo {author}
  {\bibfnamefont {Y.}~\bibnamefont {Hu}}, \bibinfo {author} {\bibfnamefont
  {X.-P.}\ \bibnamefont {Liu}}, \bibinfo {author} {\bibfnamefont {X.-M.}\
  \bibnamefont {Zhai}}, \bibinfo {author} {\bibfnamefont {P.}~\bibnamefont
  {Liu}}, \bibinfo {author} {\bibfnamefont {X.}~\bibnamefont {Jiang}}, \bibinfo
  {author} {\bibfnamefont {P.}~\bibnamefont {Xu}}, \bibinfo {author}
  {\bibfnamefont {H.-N.}\ \bibnamefont {Dai}}, \bibinfo {author} {\bibfnamefont
  {Y.-A.}\ \bibnamefont {Chen}},\ and\ \bibinfo {author} {\bibfnamefont
  {J.-W.}\ \bibnamefont {Pan}},\ }\bibfield  {title} {\bibinfo {title} {A
  strontium lattice clock with both stability and uncertainty below $5\times
  10^{-18}$},\ }\href {https://doi.org/10.1088/1681-7575/ad1a4c} {\bibfield
  {journal} {\bibinfo  {journal} {Metrologia}\ }\textbf {\bibinfo {volume}
  {61}},\ \bibinfo {pages} {015006} (\bibinfo {year} {2024})}\BibitemShut
  {NoStop}%
\bibitem [{\citenamefont {Li}\ \emph {et~al.}(2023)\citenamefont {Li},
  \citenamefont {Jia}, \citenamefont {Liu}, \citenamefont {Liu}, \citenamefont
  {Wang}, \citenamefont {Kong}, \citenamefont {Li}, \citenamefont {Cui},
  \citenamefont {Dai}, \citenamefont {Chen},\ and\ \citenamefont
  {Pan}}]{Li2023}%
  \BibitemOpen
  \bibfield  {author} {\bibinfo {author} {\bibfnamefont {J.}~\bibnamefont
  {Li}}, \bibinfo {author} {\bibfnamefont {Z.-P.}\ \bibnamefont {Jia}},
  \bibinfo {author} {\bibfnamefont {P.}~\bibnamefont {Liu}}, \bibinfo {author}
  {\bibfnamefont {X.-Y.}\ \bibnamefont {Liu}}, \bibinfo {author} {\bibfnamefont
  {D.-Z.}\ \bibnamefont {Wang}}, \bibinfo {author} {\bibfnamefont {D.-Q.}\
  \bibnamefont {Kong}}, \bibinfo {author} {\bibfnamefont {S.-P.}\ \bibnamefont
  {Li}}, \bibinfo {author} {\bibfnamefont {X.-Y.}\ \bibnamefont {Cui}},
  \bibinfo {author} {\bibfnamefont {H.-N.}\ \bibnamefont {Dai}}, \bibinfo
  {author} {\bibfnamefont {Y.-A.}\ \bibnamefont {Chen}},\ and\ \bibinfo
  {author} {\bibfnamefont {J.-W.}\ \bibnamefont {Pan}},\ }\bibfield  {title}
  {\bibinfo {title} {An integrated high-flux cold atomic beam source for
  strontium},\ }\bibfield  {journal} {\bibinfo  {journal} {Review of Scientific
  Instruments}\ }\textbf {\bibinfo {volume} {94}},\ \href
  {https://doi.org/10.1063/5.0162128} {10.1063/5.0162128} (\bibinfo {year}
  {2023})\BibitemShut {NoStop}%
\bibitem [{\citenamefont {Katori}(2011)}]{Katori2011}%
  \BibitemOpen
  \bibfield  {author} {\bibinfo {author} {\bibfnamefont {H.}~\bibnamefont
  {Katori}},\ }\bibfield  {title} {\bibinfo {title} {Optical lattice clocks and
  quantum metrology},\ }\href {https://doi.org/10.1038/nphoton.2011.45}
  {\bibfield  {journal} {\bibinfo  {journal} {Nature Photonics}\ }\textbf
  {\bibinfo {volume} {5}},\ \bibinfo {pages} {203} (\bibinfo {year}
  {2011})}\BibitemShut {NoStop}%
\bibitem [{\citenamefont {Wang}\ \emph {et~al.}(2024)\citenamefont {Wang},
  \citenamefont {Yao}, \citenamefont {Shi}, \citenamefont {Yu}, \citenamefont
  {Ma},\ and\ \citenamefont {Jiang}}]{Wang2024}%
  \BibitemOpen
  \bibfield  {author} {\bibinfo {author} {\bibfnamefont {C.}~\bibnamefont
  {Wang}}, \bibinfo {author} {\bibfnamefont {Y.}~\bibnamefont {Yao}}, \bibinfo
  {author} {\bibfnamefont {H.}~\bibnamefont {Shi}}, \bibinfo {author}
  {\bibfnamefont {H.}~\bibnamefont {Yu}}, \bibinfo {author} {\bibfnamefont
  {L.}~\bibnamefont {Ma}},\ and\ \bibinfo {author} {\bibfnamefont
  {Y.}~\bibnamefont {Jiang}},\ }\bibfield  {title} {\bibinfo {title} {A {Yb}
  optical clock with a lattice power enhancement cavity},\ }\href
  {https://doi.org/10.1088/1674-1056/ad1986} {\bibfield  {journal} {\bibinfo
  {journal} {Chinese Physics B}\ }\textbf {\bibinfo {volume} {33}},\ \bibinfo
  {pages} {030601} (\bibinfo {year} {2024})}\BibitemShut {NoStop}%
\bibitem [{\citenamefont {Fasano}\ \emph {et~al.}(2021)\citenamefont {Fasano},
  \citenamefont {Chen}, \citenamefont {McGrew}, \citenamefont {Brand},
  \citenamefont {Fox},\ and\ \citenamefont {Ludlow}}]{Fasano2021}%
  \BibitemOpen
  \bibfield  {author} {\bibinfo {author} {\bibfnamefont {R.}~\bibnamefont
  {Fasano}}, \bibinfo {author} {\bibfnamefont {Y.}~\bibnamefont {Chen}},
  \bibinfo {author} {\bibfnamefont {W.}~\bibnamefont {McGrew}}, \bibinfo
  {author} {\bibfnamefont {W.}~\bibnamefont {Brand}}, \bibinfo {author}
  {\bibfnamefont {R.}~\bibnamefont {Fox}},\ and\ \bibinfo {author}
  {\bibfnamefont {A.}~\bibnamefont {Ludlow}},\ }\bibfield  {title} {\bibinfo
  {title} {Characterization and suppression of background light shifts in an
  optical lattice clock},\ }\href
  {https://doi.org/10.1103/PhysRevApplied.15.044016} {\bibfield  {journal}
  {\bibinfo  {journal} {Phys. Rev. Appl.}\ }\textbf {\bibinfo {volume} {15}},\
  \bibinfo {pages} {044016} (\bibinfo {year} {2021})}\BibitemShut {NoStop}%
\bibitem [{\citenamefont {Jia}\ \emph {et~al.}(2025)\citenamefont {Jia},
  \citenamefont {Cui}, \citenamefont {Xie}, \citenamefont {Zhang},
  \citenamefont {Niu}, \citenamefont {Liu}, \citenamefont {Zhu}, \citenamefont
  {Li},\ and\ \citenamefont {Dai}}]{Jia2025}%
  \BibitemOpen
  \bibfield  {author} {\bibinfo {author} {\bibfnamefont {Z.-P.}\ \bibnamefont
  {Jia}}, \bibinfo {author} {\bibfnamefont {X.-Y.}\ \bibnamefont {Cui}},
  \bibinfo {author} {\bibfnamefont {Y.-J.}\ \bibnamefont {Xie}}, \bibinfo
  {author} {\bibfnamefont {X.}~\bibnamefont {Zhang}}, \bibinfo {author}
  {\bibfnamefont {G.-Z.}\ \bibnamefont {Niu}}, \bibinfo {author} {\bibfnamefont
  {X.-Y.}\ \bibnamefont {Liu}}, \bibinfo {author} {\bibfnamefont {Q.-Q.}\
  \bibnamefont {Zhu}}, \bibinfo {author} {\bibfnamefont {J.}~\bibnamefont
  {Li}},\ and\ \bibinfo {author} {\bibfnamefont {H.-N.}\ \bibnamefont {Dai}},\
  }\bibfield  {title} {\bibinfo {title} {Suppressing background spectra of
  lattice lasers in strontium optical clocks},\ }\bibfield  {journal} {\bibinfo
   {journal} {Physical Review Applied}\ }\textbf {\bibinfo {volume} {23}},\
  \href {https://doi.org/10.1103/physrevapplied.23.014014}
  {10.1103/physrevapplied.23.014014} (\bibinfo {year} {2025})\BibitemShut
  {NoStop}%
\bibitem [{\citenamefont {Nemitz}\ \emph {et~al.}(2016)\citenamefont {Nemitz},
  \citenamefont {Ohkubo}, \citenamefont {Takamoto}, \citenamefont {Ushijima},
  \citenamefont {Das}, \citenamefont {Ohmae},\ and\ \citenamefont
  {Katori}}]{Nemitz2016}%
  \BibitemOpen
  \bibfield  {author} {\bibinfo {author} {\bibfnamefont {N.}~\bibnamefont
  {Nemitz}}, \bibinfo {author} {\bibfnamefont {T.}~\bibnamefont {Ohkubo}},
  \bibinfo {author} {\bibfnamefont {M.}~\bibnamefont {Takamoto}}, \bibinfo
  {author} {\bibfnamefont {I.}~\bibnamefont {Ushijima}}, \bibinfo {author}
  {\bibfnamefont {M.}~\bibnamefont {Das}}, \bibinfo {author} {\bibfnamefont
  {N.}~\bibnamefont {Ohmae}},\ and\ \bibinfo {author} {\bibfnamefont
  {H.}~\bibnamefont {Katori}},\ }\bibfield  {title} {\bibinfo {title}
  {Frequency ratio of {Yb} and {Sr} clocks with $5\times10^{-17}$ uncertainty
  at 150 seconds averaging time},\ }\href
  {https://doi.org/10.1038/nphoton.2016.20} {\bibfield  {journal} {\bibinfo
  {journal} {Nature Photonics}\ }\textbf {\bibinfo {volume} {10}},\ \bibinfo
  {pages} {258–261} (\bibinfo {year} {2016})}\BibitemShut {NoStop}%
\bibitem [{\citenamefont {Bishof}(2014)}]{phdthesis2014}%
  \BibitemOpen
  \bibfield  {author} {\bibinfo {author} {\bibfnamefont {M.~N.}\ \bibnamefont
  {Bishof}},\ }\emph {\bibinfo {title} {Understanding atomic interactions in an
  optical lattice clock and using them to study many-body physics}},\
  \href@noop {} {Ph.D. thesis},\ \bibinfo  {school} {University of Colorado}
  (\bibinfo {year} {2014})\BibitemShut {NoStop}%
\bibitem [{\citenamefont {Falke}\ \emph {et~al.}(2014)\citenamefont {Falke},
  \citenamefont {Lemke}, \citenamefont {Grebing}, \citenamefont {Lipphardt},
  \citenamefont {Weyers}, \citenamefont {Gerginov}, \citenamefont {Huntemann},
  \citenamefont {Hagemann}, \citenamefont {Al-Masoudi}, \citenamefont
  {H\"{a}fner}, \citenamefont {Vogt}, \citenamefont {Sterr},\ and\
  \citenamefont {Lisdat}}]{Falke2014}%
  \BibitemOpen
  \bibfield  {author} {\bibinfo {author} {\bibfnamefont {S.}~\bibnamefont
  {Falke}}, \bibinfo {author} {\bibfnamefont {N.}~\bibnamefont {Lemke}},
  \bibinfo {author} {\bibfnamefont {C.}~\bibnamefont {Grebing}}, \bibinfo
  {author} {\bibfnamefont {B.}~\bibnamefont {Lipphardt}}, \bibinfo {author}
  {\bibfnamefont {S.}~\bibnamefont {Weyers}}, \bibinfo {author} {\bibfnamefont
  {V.}~\bibnamefont {Gerginov}}, \bibinfo {author} {\bibfnamefont
  {N.}~\bibnamefont {Huntemann}}, \bibinfo {author} {\bibfnamefont
  {C.}~\bibnamefont {Hagemann}}, \bibinfo {author} {\bibfnamefont
  {A.}~\bibnamefont {Al-Masoudi}}, \bibinfo {author} {\bibfnamefont
  {S.}~\bibnamefont {H\"{a}fner}}, \bibinfo {author} {\bibfnamefont
  {S.}~\bibnamefont {Vogt}}, \bibinfo {author} {\bibfnamefont {U.}~\bibnamefont
  {Sterr}},\ and\ \bibinfo {author} {\bibfnamefont {C.}~\bibnamefont
  {Lisdat}},\ }\bibfield  {title} {\bibinfo {title} {A strontium lattice clock
  with $3\times10^{-17}$ inaccuracy and its frequency},\ }\href
  {https://doi.org/10.1088/1367-2630/16/7/073023} {\bibfield  {journal}
  {\bibinfo  {journal} {New Journal of Physics}\ }\textbf {\bibinfo {volume}
  {16}},\ \bibinfo {pages} {073023} (\bibinfo {year} {2014})}\BibitemShut
  {NoStop}%
\bibitem [{\citenamefont {Oelker}\ \emph {et~al.}(2019)\citenamefont {Oelker},
  \citenamefont {Hutson}, \citenamefont {Kennedy}, \citenamefont {Sonderhouse},
  \citenamefont {Bothwell}, \citenamefont {Goban}, \citenamefont {Kedar},
  \citenamefont {Sanner}, \citenamefont {Robinson}, \citenamefont {Marti},
  \citenamefont {Matei}, \citenamefont {Legero}, \citenamefont {Giunta},
  \citenamefont {Holzwarth}, \citenamefont {Riehle}, \citenamefont {Sterr},\
  and\ \citenamefont {Ye}}]{Oelker2019}%
  \BibitemOpen
  \bibfield  {author} {\bibinfo {author} {\bibfnamefont {E.}~\bibnamefont
  {Oelker}}, \bibinfo {author} {\bibfnamefont {R.~B.}\ \bibnamefont {Hutson}},
  \bibinfo {author} {\bibfnamefont {C.~J.}\ \bibnamefont {Kennedy}}, \bibinfo
  {author} {\bibfnamefont {L.}~\bibnamefont {Sonderhouse}}, \bibinfo {author}
  {\bibfnamefont {T.}~\bibnamefont {Bothwell}}, \bibinfo {author}
  {\bibfnamefont {A.}~\bibnamefont {Goban}}, \bibinfo {author} {\bibfnamefont
  {D.}~\bibnamefont {Kedar}}, \bibinfo {author} {\bibfnamefont
  {C.}~\bibnamefont {Sanner}}, \bibinfo {author} {\bibfnamefont {J.~M.}\
  \bibnamefont {Robinson}}, \bibinfo {author} {\bibfnamefont {G.~E.}\
  \bibnamefont {Marti}}, \bibinfo {author} {\bibfnamefont {D.~G.}\ \bibnamefont
  {Matei}}, \bibinfo {author} {\bibfnamefont {T.}~\bibnamefont {Legero}},
  \bibinfo {author} {\bibfnamefont {M.}~\bibnamefont {Giunta}}, \bibinfo
  {author} {\bibfnamefont {R.}~\bibnamefont {Holzwarth}}, \bibinfo {author}
  {\bibfnamefont {F.}~\bibnamefont {Riehle}}, \bibinfo {author} {\bibfnamefont
  {U.}~\bibnamefont {Sterr}},\ and\ \bibinfo {author} {\bibfnamefont
  {J.}~\bibnamefont {Ye}},\ }\bibfield  {title} {\bibinfo {title}
  {Demonstration of $4.8\times10^{-17}$ stability at 1 s for two independent
  optical clocks},\ }\href {https://doi.org/10.1038/s41566-019-0493-4}
  {\bibfield  {journal} {\bibinfo  {journal} {Nature Photonics}\ }\textbf
  {\bibinfo {volume} {13}},\ \bibinfo {pages} {714} (\bibinfo {year}
  {2019})}\BibitemShut {NoStop}%
\bibitem [{\citenamefont {Zhu}\ \emph {et~al.}(2024)\citenamefont {Zhu},
  \citenamefont {Cui}, \citenamefont {Kong}, \citenamefont {Yu}, \citenamefont
  {Zhai}, \citenamefont {Zheng}, \citenamefont {Xie}, \citenamefont {Zhang},
  \citenamefont {Jiang}, \citenamefont {Zhang}, \citenamefont {Xu},
  \citenamefont {Dai}, \citenamefont {Chen},\ and\ \citenamefont
  {Pan}}]{Zhu2024}%
  \BibitemOpen
  \bibfield  {author} {\bibinfo {author} {\bibfnamefont {X.-Q.}\ \bibnamefont
  {Zhu}}, \bibinfo {author} {\bibfnamefont {X.-Y.}\ \bibnamefont {Cui}},
  \bibinfo {author} {\bibfnamefont {D.-Q.}\ \bibnamefont {Kong}}, \bibinfo
  {author} {\bibfnamefont {H.-W.}\ \bibnamefont {Yu}}, \bibinfo {author}
  {\bibfnamefont {X.-M.}\ \bibnamefont {Zhai}}, \bibinfo {author}
  {\bibfnamefont {M.-Y.}\ \bibnamefont {Zheng}}, \bibinfo {author}
  {\bibfnamefont {X.-P.}\ \bibnamefont {Xie}}, \bibinfo {author} {\bibfnamefont
  {Q.}~\bibnamefont {Zhang}}, \bibinfo {author} {\bibfnamefont
  {X.}~\bibnamefont {Jiang}}, \bibinfo {author} {\bibfnamefont {X.-B.}\
  \bibnamefont {Zhang}}, \bibinfo {author} {\bibfnamefont {P.}~\bibnamefont
  {Xu}}, \bibinfo {author} {\bibfnamefont {H.-N.}\ \bibnamefont {Dai}},
  \bibinfo {author} {\bibfnamefont {Y.-A.}\ \bibnamefont {Chen}},\ and\
  \bibinfo {author} {\bibfnamefont {J.-W.}\ \bibnamefont {Pan}},\ }\bibfield
  {title} {\bibinfo {title} {An ultrastable 1397-nm laser stabilized by a
  crystalline-coated room-temperature cavity},\ }\bibfield  {journal} {\bibinfo
   {journal} {Review of Scientific Instruments}\ }\textbf {\bibinfo {volume}
  {95}},\ \href {https://doi.org/10.1063/5.0200553} {10.1063/5.0200553}
  (\bibinfo {year} {2024})\BibitemShut {NoStop}%
\bibitem [{\citenamefont {Yu}\ \emph {et~al.}(2025)\citenamefont {Yu},
  \citenamefont {Liu}, \citenamefont {Li}, \citenamefont {Jia}, \citenamefont
  {Zhang}, \citenamefont {Yan}, \citenamefont {Li}, \citenamefont {Dai},\ and\
  \citenamefont {Chen}}]{Yu2025}%
  \BibitemOpen
  \bibfield  {author} {\bibinfo {author} {\bibfnamefont {H.}~\bibnamefont
  {Yu}}, \bibinfo {author} {\bibfnamefont {P.}~\bibnamefont {Liu}}, \bibinfo
  {author} {\bibfnamefont {Y.}~\bibnamefont {Li}}, \bibinfo {author}
  {\bibfnamefont {Z.}~\bibnamefont {Jia}}, \bibinfo {author} {\bibfnamefont
  {X.}~\bibnamefont {Zhang}}, \bibinfo {author} {\bibfnamefont
  {J.}~\bibnamefont {Yan}}, \bibinfo {author} {\bibfnamefont {J.}~\bibnamefont
  {Li}}, \bibinfo {author} {\bibfnamefont {H.}~\bibnamefont {Dai}},\ and\
  \bibinfo {author} {\bibfnamefont {Y.}~\bibnamefont {Chen}},\ }\bibfield
  {title} {\bibinfo {title} {Improved evaluation of blackbody radiation shift
  with uncertainty below $1\times10^{-18}$ in a strontium lattice clock},\
  }\href {https://doi.org/https://doi.org/10.1016/j.measurement.2025.118527}
  {\bibfield  {journal} {\bibinfo  {journal} {Measurement}\ ,\ \bibinfo {pages}
  {118527}} (\bibinfo {year} {2025})}\BibitemShut {NoStop}%
\bibitem [{\citenamefont {Dole{\v{z}}al}\ \emph {et~al.}(2015)\citenamefont
  {Dole{\v{z}}al}, \citenamefont {Balling}, \citenamefont {Nisbet-Jones},
  \citenamefont {King}, \citenamefont {Jones}, \citenamefont {Klein},
  \citenamefont {Gill}, \citenamefont {Lindvall}, \citenamefont {Wallin},
  \citenamefont {Merimaa} \emph {et~al.}}]{Dolezal2015}%
  \BibitemOpen
  \bibfield  {author} {\bibinfo {author} {\bibfnamefont {M.}~\bibnamefont
  {Dole{\v{z}}al}}, \bibinfo {author} {\bibfnamefont {P.}~\bibnamefont
  {Balling}}, \bibinfo {author} {\bibfnamefont {P.~B.}\ \bibnamefont
  {Nisbet-Jones}}, \bibinfo {author} {\bibfnamefont {S.~A.}\ \bibnamefont
  {King}}, \bibinfo {author} {\bibfnamefont {J.~M.}\ \bibnamefont {Jones}},
  \bibinfo {author} {\bibfnamefont {H.~A.}\ \bibnamefont {Klein}}, \bibinfo
  {author} {\bibfnamefont {P.}~\bibnamefont {Gill}}, \bibinfo {author}
  {\bibfnamefont {T.}~\bibnamefont {Lindvall}}, \bibinfo {author}
  {\bibfnamefont {A.~E.}\ \bibnamefont {Wallin}}, \bibinfo {author}
  {\bibfnamefont {M.}~\bibnamefont {Merimaa}}, \emph {et~al.},\ }\bibfield
  {title} {\bibinfo {title} {Analysis of thermal radiation in ion traps for
  optical frequency standards},\ }\href
  {https://doi.org/10.1088/0026-1394/52/6/842} {\bibfield  {journal} {\bibinfo
  {journal} {Metrologia}\ }\textbf {\bibinfo {volume} {52}},\ \bibinfo {pages}
  {842} (\bibinfo {year} {2015})}\BibitemShut {NoStop}%
\bibitem [{\citenamefont {Xiong}\ \emph {et~al.}(2021)\citenamefont {Xiong},
  \citenamefont {Zhu}, \citenamefont {Wang}, \citenamefont {Zhang},
  \citenamefont {Tian}, \citenamefont {Wang}, \citenamefont {He}, \citenamefont
  {Xiong},\ and\ \citenamefont {Lyu}}]{Xiong2021}%
  \BibitemOpen
  \bibfield  {author} {\bibinfo {author} {\bibfnamefont {D.}~\bibnamefont
  {Xiong}}, \bibinfo {author} {\bibfnamefont {Q.}~\bibnamefont {Zhu}}, \bibinfo
  {author} {\bibfnamefont {J.}~\bibnamefont {Wang}}, \bibinfo {author}
  {\bibfnamefont {A.}~\bibnamefont {Zhang}}, \bibinfo {author} {\bibfnamefont
  {C.}~\bibnamefont {Tian}}, \bibinfo {author} {\bibfnamefont {B.}~\bibnamefont
  {Wang}}, \bibinfo {author} {\bibfnamefont {L.}~\bibnamefont {He}}, \bibinfo
  {author} {\bibfnamefont {Z.}~\bibnamefont {Xiong}},\ and\ \bibinfo {author}
  {\bibfnamefont {B.}~\bibnamefont {Lyu}},\ }\bibfield  {title} {\bibinfo
  {title} {Finite element analysis of blackbody radiation environment for an
  ytterbium lattice clock operated at room temperature},\ }\href
  {https://doi.org/10.1088/1681-7575/abeec3} {\bibfield  {journal} {\bibinfo
  {journal} {Metrologia}\ }\textbf {\bibinfo {volume} {58}},\ \bibinfo {pages}
  {035005} (\bibinfo {year} {2021})}\BibitemShut {NoStop}%
\bibitem [{\citenamefont {Middelmann}\ \emph {et~al.}(2012)\citenamefont
  {Middelmann}, \citenamefont {Falke}, \citenamefont {Lisdat},\ and\
  \citenamefont {Sterr}}]{Middelmann2012}%
  \BibitemOpen
  \bibfield  {author} {\bibinfo {author} {\bibfnamefont {T.}~\bibnamefont
  {Middelmann}}, \bibinfo {author} {\bibfnamefont {S.}~\bibnamefont {Falke}},
  \bibinfo {author} {\bibfnamefont {C.}~\bibnamefont {Lisdat}},\ and\ \bibinfo
  {author} {\bibfnamefont {U.}~\bibnamefont {Sterr}},\ }\bibfield  {title}
  {\bibinfo {title} {High accuracy correction of blackbody radiation shift in
  an optical lattice clock},\ }\href
  {https://doi.org/https://doi.org/10.1103/PhysRevLett.109.263004} {\bibfield
  {journal} {\bibinfo  {journal} {Physical Review Letters}\ }\textbf {\bibinfo
  {volume} {109}},\ \bibinfo {pages} {263004} (\bibinfo {year}
  {2012})}\BibitemShut {NoStop}%
\bibitem [{\citenamefont {Bishof}\ \emph {et~al.}(2011)\citenamefont {Bishof},
  \citenamefont {Martin}, \citenamefont {Swallows}, \citenamefont {Benko},
  \citenamefont {Lin}, \citenamefont {Quéméner}, \citenamefont {Rey},\ and\
  \citenamefont {Ye}}]{Bishof2011}%
  \BibitemOpen
  \bibfield  {author} {\bibinfo {author} {\bibfnamefont {M.}~\bibnamefont
  {Bishof}}, \bibinfo {author} {\bibfnamefont {M.~J.}\ \bibnamefont {Martin}},
  \bibinfo {author} {\bibfnamefont {M.~D.}\ \bibnamefont {Swallows}}, \bibinfo
  {author} {\bibfnamefont {C.}~\bibnamefont {Benko}}, \bibinfo {author}
  {\bibfnamefont {Y.}~\bibnamefont {Lin}}, \bibinfo {author} {\bibfnamefont
  {G.}~\bibnamefont {Quéméner}}, \bibinfo {author} {\bibfnamefont {A.~M.}\
  \bibnamefont {Rey}},\ and\ \bibinfo {author} {\bibfnamefont {J.}~\bibnamefont
  {Ye}},\ }\bibfield  {title} {\bibinfo {title} {Inelastic collisions and
  density-dependent excitation suppression in a ${}^{87}${Sr} optical lattice
  clock},\ }\bibfield  {journal} {\bibinfo  {journal} {Physical Review A}\
  }\textbf {\bibinfo {volume} {84}},\ \href
  {https://doi.org/10.1103/physreva.84.052716} {10.1103/physreva.84.052716}
  (\bibinfo {year} {2011})\BibitemShut {NoStop}%
\bibitem [{\citenamefont {Swallows}\ \emph {et~al.}(2012)\citenamefont
  {Swallows}, \citenamefont {Martin}, \citenamefont {Bishof}, \citenamefont
  {Benko}, \citenamefont {Lin}, \citenamefont {Blatt}, \citenamefont {Rey},\
  and\ \citenamefont {Ye}}]{Swallows2012}%
  \BibitemOpen
  \bibfield  {author} {\bibinfo {author} {\bibfnamefont {M.~D.}\ \bibnamefont
  {Swallows}}, \bibinfo {author} {\bibfnamefont {M.~J.}\ \bibnamefont
  {Martin}}, \bibinfo {author} {\bibfnamefont {M.}~\bibnamefont {Bishof}},
  \bibinfo {author} {\bibfnamefont {C.}~\bibnamefont {Benko}}, \bibinfo
  {author} {\bibfnamefont {Y.}~\bibnamefont {Lin}}, \bibinfo {author}
  {\bibfnamefont {S.}~\bibnamefont {Blatt}}, \bibinfo {author} {\bibfnamefont
  {A.~M.}\ \bibnamefont {Rey}},\ and\ \bibinfo {author} {\bibfnamefont
  {J.}~\bibnamefont {Ye}},\ }\bibfield  {title} {\bibinfo {title} {Operating a
  {${}^{87}${Sr}} optical lattice clock with high precision and at high
  density},\ }\href {https://doi.org/10.1109/tuffc.2012.2210} {\bibfield
  {journal} {\bibinfo  {journal} {IEEE Transactions on Ultrasonics,
  Ferroelectrics and Frequency Control}\ }\textbf {\bibinfo {volume} {59}},\
  \bibinfo {pages} {416–425} (\bibinfo {year} {2012})}\BibitemShut {NoStop}%
\bibitem [{\citenamefont {Campbell}\ \emph {et~al.}(2009)\citenamefont
  {Campbell}, \citenamefont {Boyd}, \citenamefont {Thomsen}, \citenamefont
  {Martin}, \citenamefont {Blatt}, \citenamefont {Swallows}, \citenamefont
  {Nicholson}, \citenamefont {Fortier}, \citenamefont {Oates}, \citenamefont
  {Diddams}, \citenamefont {Lemke}, \citenamefont {Naidon}, \citenamefont
  {Julienne}, \citenamefont {Ye},\ and\ \citenamefont {Ludlow}}]{Campbell2009}%
  \BibitemOpen
  \bibfield  {author} {\bibinfo {author} {\bibfnamefont {G.~K.}\ \bibnamefont
  {Campbell}}, \bibinfo {author} {\bibfnamefont {M.~M.}\ \bibnamefont {Boyd}},
  \bibinfo {author} {\bibfnamefont {J.~W.}\ \bibnamefont {Thomsen}}, \bibinfo
  {author} {\bibfnamefont {M.~J.}\ \bibnamefont {Martin}}, \bibinfo {author}
  {\bibfnamefont {S.}~\bibnamefont {Blatt}}, \bibinfo {author} {\bibfnamefont
  {M.~D.}\ \bibnamefont {Swallows}}, \bibinfo {author} {\bibfnamefont {T.~L.}\
  \bibnamefont {Nicholson}}, \bibinfo {author} {\bibfnamefont {T.}~\bibnamefont
  {Fortier}}, \bibinfo {author} {\bibfnamefont {C.~W.}\ \bibnamefont {Oates}},
  \bibinfo {author} {\bibfnamefont {S.~A.}\ \bibnamefont {Diddams}}, \bibinfo
  {author} {\bibfnamefont {N.~D.}\ \bibnamefont {Lemke}}, \bibinfo {author}
  {\bibfnamefont {P.}~\bibnamefont {Naidon}}, \bibinfo {author} {\bibfnamefont
  {P.}~\bibnamefont {Julienne}}, \bibinfo {author} {\bibfnamefont
  {J.}~\bibnamefont {Ye}},\ and\ \bibinfo {author} {\bibfnamefont {A.~D.}\
  \bibnamefont {Ludlow}},\ }\bibfield  {title} {\bibinfo {title} {Probing
  interactions between ultracold fermions},\ }\href
  {https://doi.org/10.1126/science.1169724} {\bibfield  {journal} {\bibinfo
  {journal} {Science}\ }\textbf {\bibinfo {volume} {324}},\ \bibinfo {pages}
  {360–363} (\bibinfo {year} {2009})}\BibitemShut {NoStop}%
\bibitem [{\citenamefont {Swallows}\ \emph {et~al.}(2011)\citenamefont
  {Swallows}, \citenamefont {Bishof}, \citenamefont {Lin}, \citenamefont
  {Blatt}, \citenamefont {Martin}, \citenamefont {Rey},\ and\ \citenamefont
  {Ye}}]{Swallows2011}%
  \BibitemOpen
  \bibfield  {author} {\bibinfo {author} {\bibfnamefont {M.~D.}\ \bibnamefont
  {Swallows}}, \bibinfo {author} {\bibfnamefont {M.}~\bibnamefont {Bishof}},
  \bibinfo {author} {\bibfnamefont {Y.}~\bibnamefont {Lin}}, \bibinfo {author}
  {\bibfnamefont {S.}~\bibnamefont {Blatt}}, \bibinfo {author} {\bibfnamefont
  {M.~J.}\ \bibnamefont {Martin}}, \bibinfo {author} {\bibfnamefont {A.~M.}\
  \bibnamefont {Rey}},\ and\ \bibinfo {author} {\bibfnamefont {J.}~\bibnamefont
  {Ye}},\ }\bibfield  {title} {\bibinfo {title} {Suppression of collisional
  shifts in a strongly interacting lattice clock},\ }\href
  {https://doi.org/10.1126/science.1196442} {\bibfield  {journal} {\bibinfo
  {journal} {Science}\ }\textbf {\bibinfo {volume} {331}},\ \bibinfo {pages}
  {1043–1046} (\bibinfo {year} {2011})}\BibitemShut {NoStop}%
\bibitem [{\citenamefont {Lemke}\ \emph {et~al.}(2011)\citenamefont {Lemke},
  \citenamefont {von Stecher}, \citenamefont {Sherman}, \citenamefont {Rey},
  \citenamefont {Oates},\ and\ \citenamefont {Ludlow}}]{Lemke2011}%
  \BibitemOpen
  \bibfield  {author} {\bibinfo {author} {\bibfnamefont {N.~D.}\ \bibnamefont
  {Lemke}}, \bibinfo {author} {\bibfnamefont {J.}~\bibnamefont {von Stecher}},
  \bibinfo {author} {\bibfnamefont {J.~A.}\ \bibnamefont {Sherman}}, \bibinfo
  {author} {\bibfnamefont {A.~M.}\ \bibnamefont {Rey}}, \bibinfo {author}
  {\bibfnamefont {C.~W.}\ \bibnamefont {Oates}},\ and\ \bibinfo {author}
  {\bibfnamefont {A.~D.}\ \bibnamefont {Ludlow}},\ }\bibfield  {title}
  {\bibinfo {title} {$p$-wave cold collisions in an optical lattice clock},\
  }\bibfield  {journal} {\bibinfo  {journal} {Physical Review Letters}\
  }\textbf {\bibinfo {volume} {107}},\ \href
  {https://doi.org/10.1103/physrevlett.107.103902}
  {10.1103/physrevlett.107.103902} (\bibinfo {year} {2011})\BibitemShut
  {NoStop}%
\bibitem [{\citenamefont {Zhou}\ \emph {et~al.}(2023)\citenamefont {Zhou},
  \citenamefont {Zhang},\ and\ \citenamefont {Wang}}]{Zhou2023}%
  \BibitemOpen
  \bibfield  {author} {\bibinfo {author} {\bibfnamefont {Y.-H.}\ \bibnamefont
  {Zhou}}, \bibinfo {author} {\bibfnamefont {X.-F.}\ \bibnamefont {Zhang}},\
  and\ \bibinfo {author} {\bibfnamefont {T.}~\bibnamefont {Wang}},\ }\bibfield
  {title} {\bibinfo {title} {Density shift of optical lattice clocks via the
  multiband sampling exact diagonalization method},\ }\bibfield  {journal}
  {\bibinfo  {journal} {Physical Review A}\ }\textbf {\bibinfo {volume}
  {108}},\ \href {https://doi.org/10.1103/physreva.108.033304}
  {10.1103/physreva.108.033304} (\bibinfo {year} {2023})\BibitemShut {NoStop}%
\bibitem [{\citenamefont {Ludlow}\ \emph {et~al.}(2011)\citenamefont {Ludlow},
  \citenamefont {Lemke}, \citenamefont {Sherman}, \citenamefont {Oates},
  \citenamefont {Quéméner}, \citenamefont {von Stecher},\ and\ \citenamefont
  {Rey}}]{Ludlow2011}%
  \BibitemOpen
  \bibfield  {author} {\bibinfo {author} {\bibfnamefont {A.~D.}\ \bibnamefont
  {Ludlow}}, \bibinfo {author} {\bibfnamefont {N.~D.}\ \bibnamefont {Lemke}},
  \bibinfo {author} {\bibfnamefont {J.~A.}\ \bibnamefont {Sherman}}, \bibinfo
  {author} {\bibfnamefont {C.~W.}\ \bibnamefont {Oates}}, \bibinfo {author}
  {\bibfnamefont {G.}~\bibnamefont {Quéméner}}, \bibinfo {author}
  {\bibfnamefont {J.}~\bibnamefont {von Stecher}},\ and\ \bibinfo {author}
  {\bibfnamefont {A.~M.}\ \bibnamefont {Rey}},\ }\bibfield  {title} {\bibinfo
  {title} {Cold-collision-shift cancellation and inelastic scattering in a {Yb}
  optical lattice clock},\ }\bibfield  {journal} {\bibinfo  {journal} {Physical
  Review A}\ }\textbf {\bibinfo {volume} {84}},\ \href
  {https://doi.org/10.1103/physreva.84.052724} {10.1103/physreva.84.052724}
  (\bibinfo {year} {2011})\BibitemShut {NoStop}%
\bibitem [{\citenamefont {Lu}\ \emph {et~al.}(2023)\citenamefont {Lu},
  \citenamefont {Guo}, \citenamefont {Wang}, \citenamefont {Xu}, \citenamefont
  {Zhou}, \citenamefont {Xia}, \citenamefont {Wu},\ and\ \citenamefont
  {Chang}}]{Lu2023}%
  \BibitemOpen
  \bibfield  {author} {\bibinfo {author} {\bibfnamefont {X.}~\bibnamefont
  {Lu}}, \bibinfo {author} {\bibfnamefont {F.}~\bibnamefont {Guo}}, \bibinfo
  {author} {\bibfnamefont {Y.}~\bibnamefont {Wang}}, \bibinfo {author}
  {\bibfnamefont {Q.}~\bibnamefont {Xu}}, \bibinfo {author} {\bibfnamefont
  {C.}~\bibnamefont {Zhou}}, \bibinfo {author} {\bibfnamefont {J.}~\bibnamefont
  {Xia}}, \bibinfo {author} {\bibfnamefont {W.}~\bibnamefont {Wu}},\ and\
  \bibinfo {author} {\bibfnamefont {H.}~\bibnamefont {Chang}},\ }\bibfield
  {title} {\bibinfo {title} {Absolute frequency measurement of the
  ${}^{87}${Sr} optical lattice clock at {NTSC} using international atomic
  time},\ }\href {https://doi.org/10.1088/1681-7575/acb05c} {\bibfield
  {journal} {\bibinfo  {journal} {Metrologia}\ }\textbf {\bibinfo {volume}
  {60}},\ \bibinfo {pages} {015008} (\bibinfo {year} {2023})}\BibitemShut
  {NoStop}%
\bibitem [{\citenamefont {Katori}\ \emph {et~al.}(2003)\citenamefont {Katori},
  \citenamefont {Takamoto}, \citenamefont {Pal'chikov},\ and\ \citenamefont
  {Ovsiannikov}}]{Katori2003}%
  \BibitemOpen
  \bibfield  {author} {\bibinfo {author} {\bibfnamefont {H.}~\bibnamefont
  {Katori}}, \bibinfo {author} {\bibfnamefont {M.}~\bibnamefont {Takamoto}},
  \bibinfo {author} {\bibfnamefont {V.~G.}\ \bibnamefont {Pal'chikov}},\ and\
  \bibinfo {author} {\bibfnamefont {V.~D.}\ \bibnamefont {Ovsiannikov}},\
  }\bibfield  {title} {\bibinfo {title} {Ultrastable optical clock with neutral
  atoms in an engineered light shift trap},\ }\href
  {https://doi.org/10.1103/PhysRevLett.91.173005} {\bibfield  {journal}
  {\bibinfo  {journal} {Phys. Rev. Lett.}\ }\textbf {\bibinfo {volume} {91}},\
  \bibinfo {pages} {173005} (\bibinfo {year} {2003})}\BibitemShut {NoStop}%
\bibitem [{\citenamefont {Takamoto}\ \emph {et~al.}(2005)\citenamefont
  {Takamoto}, \citenamefont {Hong}, \citenamefont {Higashi},\ and\
  \citenamefont {Katori}}]{Takamoto2005}%
  \BibitemOpen
  \bibfield  {author} {\bibinfo {author} {\bibfnamefont {M.}~\bibnamefont
  {Takamoto}}, \bibinfo {author} {\bibfnamefont {F.-L.}\ \bibnamefont {Hong}},
  \bibinfo {author} {\bibfnamefont {R.}~\bibnamefont {Higashi}},\ and\ \bibinfo
  {author} {\bibfnamefont {H.}~\bibnamefont {Katori}},\ }\bibfield  {title}
  {\bibinfo {title} {An optical lattice clock},\ }\href
  {https://doi.org/10.1038/nature03541} {\bibfield  {journal} {\bibinfo
  {journal} {Nature}\ }\textbf {\bibinfo {volume} {435}},\ \bibinfo {pages}
  {321} (\bibinfo {year} {2005})}\BibitemShut {NoStop}%
\bibitem [{\citenamefont {Bothwell}\ \emph {et~al.}(2025)\citenamefont
  {Bothwell}, \citenamefont {Hunt}, \citenamefont {Siegel}, \citenamefont
  {Hassan}, \citenamefont {Grogan}, \citenamefont {Kobayashi}, \citenamefont
  {Gibble}, \citenamefont {Porsev}, \citenamefont {Safronova}, \citenamefont
  {Brown}, \citenamefont {Beloy},\ and\ \citenamefont {Ludlow}}]{Bothwell2025}%
  \BibitemOpen
  \bibfield  {author} {\bibinfo {author} {\bibfnamefont {T.}~\bibnamefont
  {Bothwell}}, \bibinfo {author} {\bibfnamefont {B.~D.}\ \bibnamefont {Hunt}},
  \bibinfo {author} {\bibfnamefont {J.~L.}\ \bibnamefont {Siegel}}, \bibinfo
  {author} {\bibfnamefont {Y.~S.}\ \bibnamefont {Hassan}}, \bibinfo {author}
  {\bibfnamefont {T.}~\bibnamefont {Grogan}}, \bibinfo {author} {\bibfnamefont
  {T.}~\bibnamefont {Kobayashi}}, \bibinfo {author} {\bibfnamefont
  {K.}~\bibnamefont {Gibble}}, \bibinfo {author} {\bibfnamefont {S.~G.}\
  \bibnamefont {Porsev}}, \bibinfo {author} {\bibfnamefont {M.~S.}\
  \bibnamefont {Safronova}}, \bibinfo {author} {\bibfnamefont {R.~C.}\
  \bibnamefont {Brown}}, \bibinfo {author} {\bibfnamefont {K.}~\bibnamefont
  {Beloy}},\ and\ \bibinfo {author} {\bibfnamefont {A.~D.}\ \bibnamefont
  {Ludlow}},\ }\bibfield  {title} {\bibinfo {title} {Lattice light shift
  evaluations in a dual-ensemble {Yb} optical lattice clock},\ }\bibfield
  {journal} {\bibinfo  {journal} {Physical Review Letters}\ }\textbf {\bibinfo
  {volume} {134}},\ \href {https://doi.org/10.1103/physrevlett.134.033201}
  {10.1103/physrevlett.134.033201} (\bibinfo {year} {2025})\BibitemShut
  {NoStop}%
\bibitem [{\citenamefont {Shi}\ \emph {et~al.}(2015)\citenamefont {Shi},
  \citenamefont {Robyr}, \citenamefont {Eismann}, \citenamefont {Zawada},
  \citenamefont {Lorini}, \citenamefont {Le~Targat},\ and\ \citenamefont
  {Lodewyck}}]{Shi2015}%
  \BibitemOpen
  \bibfield  {author} {\bibinfo {author} {\bibfnamefont {C.}~\bibnamefont
  {Shi}}, \bibinfo {author} {\bibfnamefont {J.-L.}\ \bibnamefont {Robyr}},
  \bibinfo {author} {\bibfnamefont {U.}~\bibnamefont {Eismann}}, \bibinfo
  {author} {\bibfnamefont {M.}~\bibnamefont {Zawada}}, \bibinfo {author}
  {\bibfnamefont {L.}~\bibnamefont {Lorini}}, \bibinfo {author} {\bibfnamefont
  {R.}~\bibnamefont {Le~Targat}},\ and\ \bibinfo {author} {\bibfnamefont
  {J.}~\bibnamefont {Lodewyck}},\ }\bibfield  {title} {\bibinfo {title}
  {Polarizabilities of the $^{87}${Sr} clock transition},\ }\href
  {https://doi.org/10.1103/PhysRevA.92.012516} {\bibfield  {journal} {\bibinfo
  {journal} {Phys. Rev. A}\ }\textbf {\bibinfo {volume} {92}},\ \bibinfo
  {pages} {012516} (\bibinfo {year} {2015})}\BibitemShut {NoStop}%
\bibitem [{\citenamefont {D\"{o}rscher}\ \emph {et~al.}(2023)\citenamefont
  {D\"{o}rscher}, \citenamefont {Klose}, \citenamefont {Maratha~Palli},\ and\
  \citenamefont {Lisdat}}]{Drscher2023}%
  \BibitemOpen
  \bibfield  {author} {\bibinfo {author} {\bibfnamefont {S.}~\bibnamefont
  {D\"{o}rscher}}, \bibinfo {author} {\bibfnamefont {J.}~\bibnamefont {Klose}},
  \bibinfo {author} {\bibfnamefont {S.}~\bibnamefont {Maratha~Palli}},\ and\
  \bibinfo {author} {\bibfnamefont {C.}~\bibnamefont {Lisdat}},\ }\bibfield
  {title} {\bibinfo {title} {Experimental determination of the {$E2\text{-}M1$}
  polarizability of the strontium clock transition},\ }\bibfield  {journal}
  {\bibinfo  {journal} {Physical Review Research}\ }\textbf {\bibinfo {volume}
  {5}},\ \href {https://doi.org/10.1103/physrevresearch.5.l012013}
  {10.1103/physrevresearch.5.l012013} (\bibinfo {year} {2023})\BibitemShut
  {NoStop}%
\bibitem [{\citenamefont {Wu}\ \emph {et~al.}(2023)\citenamefont {Wu},
  \citenamefont {Shi}, \citenamefont {Ni},\ and\ \citenamefont
  {Tang}}]{Wu2023}%
  \BibitemOpen
  \bibfield  {author} {\bibinfo {author} {\bibfnamefont {F.-F.}\ \bibnamefont
  {Wu}}, \bibinfo {author} {\bibfnamefont {T.-Y.}\ \bibnamefont {Shi}},
  \bibinfo {author} {\bibfnamefont {W.-T.}\ \bibnamefont {Ni}},\ and\ \bibinfo
  {author} {\bibfnamefont {L.-Y.}\ \bibnamefont {Tang}},\ }\bibfield  {title}
  {\bibinfo {title} {Contribution of negative-energy states to the
  {$E2\text{-}M1$} polarizability of optical clocks},\ }\bibfield  {journal}
  {\bibinfo  {journal} {Physical Review A}\ }\textbf {\bibinfo {volume}
  {108}},\ \href {https://doi.org/10.1103/physreva.108.l051101}
  {10.1103/physreva.108.l051101} (\bibinfo {year} {2023})\BibitemShut {NoStop}%
\bibitem [{\citenamefont {Lodewyck}\ \emph {et~al.}(2012)\citenamefont
  {Lodewyck}, \citenamefont {Zawada}, \citenamefont {Lorini}, \citenamefont
  {Gurov},\ and\ \citenamefont {Lemonde}}]{Lodewyck2012}%
  \BibitemOpen
  \bibfield  {author} {\bibinfo {author} {\bibfnamefont {J.}~\bibnamefont
  {Lodewyck}}, \bibinfo {author} {\bibfnamefont {M.}~\bibnamefont {Zawada}},
  \bibinfo {author} {\bibfnamefont {L.}~\bibnamefont {Lorini}}, \bibinfo
  {author} {\bibfnamefont {M.}~\bibnamefont {Gurov}},\ and\ \bibinfo {author}
  {\bibfnamefont {P.}~\bibnamefont {Lemonde}},\ }\bibfield  {title} {\bibinfo
  {title} {Observation and cancellation of a perturbing dc stark shift in
  strontium optical lattice clocks},\ }\href
  {https://doi.org/10.1109/tuffc.2012.2209} {\bibfield  {journal} {\bibinfo
  {journal} {IEEE Transactions on Ultrasonics, Ferroelectrics and Frequency
  Control}\ }\textbf {\bibinfo {volume} {59}},\ \bibinfo {pages} {411–415}
  (\bibinfo {year} {2012})}\BibitemShut {NoStop}%
\bibitem [{\citenamefont {Alves}\ \emph {et~al.}(2019)\citenamefont {Alves},
  \citenamefont {Foucault}, \citenamefont {Vallet},\ and\ \citenamefont
  {Lodewyck}}]{Alves2019}%
  \BibitemOpen
  \bibfield  {author} {\bibinfo {author} {\bibfnamefont {B.~X.~R.}\
  \bibnamefont {Alves}}, \bibinfo {author} {\bibfnamefont {Y.}~\bibnamefont
  {Foucault}}, \bibinfo {author} {\bibfnamefont {G.}~\bibnamefont {Vallet}},\
  and\ \bibinfo {author} {\bibfnamefont {J.}~\bibnamefont {Lodewyck}},\
  }\bibfield  {title} {\bibinfo {title} {Background gas collision frequency
  shift on lattice-trapped strontium atoms},\ }in\ \href
  {https://doi.org/10.1109/fcs.2019.8856042} {\emph {\bibinfo {booktitle} {2019
  Joint Conference of the IEEE International Frequency Control Symposium and
  European Frequency and Time Forum (EFTF/IFC)}}}\ (\bibinfo  {publisher}
  {IEEE},\ \bibinfo {year} {2019})\ p.\ \bibinfo {pages} {1–2}\BibitemShut
  {NoStop}%
\bibitem [{\citenamefont {Lemonde}\ and\ \citenamefont
  {Wolf}(2005)}]{Lemonde2005}%
  \BibitemOpen
  \bibfield  {author} {\bibinfo {author} {\bibfnamefont {P.}~\bibnamefont
  {Lemonde}}\ and\ \bibinfo {author} {\bibfnamefont {P.}~\bibnamefont {Wolf}},\
  }\bibfield  {title} {\bibinfo {title} {Optical lattice clock with atoms
  confined in a shallow trap},\ }\bibfield  {journal} {\bibinfo  {journal}
  {Physical Review A}\ }\textbf {\bibinfo {volume} {72}},\ \href
  {https://doi.org/10.1103/physreva.72.033409} {10.1103/physreva.72.033409}
  (\bibinfo {year} {2005})\BibitemShut {NoStop}%
\bibitem [{\citenamefont {Xu}\ \emph {et~al.}(2021)\citenamefont {Xu},
  \citenamefont {Lu}, \citenamefont {Xia}, \citenamefont {Wang},\ and\
  \citenamefont {Chang}}]{Xu2021}%
  \BibitemOpen
  \bibfield  {author} {\bibinfo {author} {\bibfnamefont {Q.}~\bibnamefont
  {Xu}}, \bibinfo {author} {\bibfnamefont {X.}~\bibnamefont {Lu}}, \bibinfo
  {author} {\bibfnamefont {J.}~\bibnamefont {Xia}}, \bibinfo {author}
  {\bibfnamefont {Y.}~\bibnamefont {Wang}},\ and\ \bibinfo {author}
  {\bibfnamefont {H.}~\bibnamefont {Chang}},\ }\bibfield  {title} {\bibinfo
  {title} {Measuring the probe stark shift by frequency modulation spectroscopy
  in an ${}^{87}${Sr} optical lattice clock},\ }\bibfield  {journal} {\bibinfo
  {journal} {Applied Physics Letters}\ }\textbf {\bibinfo {volume} {119}},\
  \href {https://doi.org/10.1063/5.0060277} {10.1063/5.0060277} (\bibinfo
  {year} {2021})\BibitemShut {NoStop}%
\end{thebibliography}%

\end{document}